\documentclass[a4paper,preprintnumbers,showpacs,onecolumn,superscriptaddress,nofootinbib,amsmath,amssymb,notitlepage]{revtex4-2}

\usepackage{mathtools,leftindex,tensor,mhchem}
\usepackage{booktabs}
\usepackage{siunitx}
\usepackage{enumitem}
\usepackage{times}
\usepackage{dcolumn}
\usepackage{mathrsfs}
\usepackage{comment}
\usepackage{hyperref}
\usepackage{amsmath,accents}
\usepackage{mathtools}
\usepackage{bbm}
\usepackage{bm}
\usepackage{amsfonts}
\usepackage{amssymb}
\usepackage{mathrsfs}
\usepackage[scr=rsfs,cal=boondox]{mathalfa}
\usepackage{wasysym}
\usepackage{graphicx}
\usepackage[]{subfigure}
\usepackage{filecontents}
\usepackage[dvipsnames]{xcolor}
\usepackage{bbold}
\usepackage[scr=rsfs,cal=boondox]{mathalfa}

\usepackage{stackengine}
\stackMath
\newcommand\tsup[2][2]{%
 \def\useanchorwidth{T}%
  \ifnum#1>1%
    \stackon[-1.3ex]{\tsup[\numexpr#1-1\relax]{#2}}{\mathchar"307E}%
  \else%
    \stackon[-1ex]{#2}{\mathchar"307E}%
  \fi%
}
\hypersetup{
    bookmarks=true,         
    pdftoolbar=true,        
    pdfmenubar=true,        
    pdffitwindow=true,     
    pdfstartview={FitH},     
    pdftitle={My title},     
    pdfauthor={author},      
    pdfsubject={Subject},    
    pdfcreator={Creator},    
    pdfproducer={Producer},  
    pdfkeywords={keyword1} 
    pdfnewwindow=true,       
    colorlinks=true ,        
    linkcolor=Blue  ,        
    citecolor=Blue,      
    filecolor=Blue,       
    urlcolor=Blue            
}

\newcommand{\ed}{\mathrm{d}}

\newcommand{\mB}{\mathcal{B}}

\newcommand{\R}{\mathcal{R}}

\newcommand{\cn}{\operatorname{\mathbf{cn}}}
\newcommand{\nc}{\operatorname{\mathbf{nc}}}

\newcommand{\m}{\mathcal{k}}

\newcommand{\Q}{\mathscr{Q}}

\newcommand{\oalpha}[1]{\accentset{\circ}{\alpha}}
\newcommand{\obf}[1]{\accentset{\circ}{\mathbf{f}}}
\newcommand{\boR}[1]{\accentset{\circ}{\mathbf{R}}}
\newcommand{\obF}[1]{\accentset{\circ}{\mathbf{F}}}
\newcommand{\obPi}[1]{\accentset{\circ}{\mathbf{\Pi}}}

\usepackage{scalerel}
\usepackage{tikz}
\usetikzlibrary{svg.path}

\definecolor{orcidlogocol}{HTML}{A6CE39}
\tikzset{
  orcidlogo/.pic={
    \fill[orcidlogocol] svg{M256,128c0,70.7-57.3,128-128,128C57.3,256,0,198.7,0,128C0,57.3,57.3,0,128,0C198.7,0,256,57.3,256,128z};
    \fill[white] svg{M86.3,186.2H70.9V79.1h15.4v48.4V186.2z}
                 svg{M108.9,79.1h41.6c39.6,0,57,28.3,57,53.6c0,27.5-21.5,53.6-56.8,53.6h-41.8V79.1z M124.3,172.4h24.5c34.9,0,42.9-26.5,42.9-39.7c0-21.5-13.7-39.7-43.7-39.7h-23.7V172.4z}
                 svg{M88.7,56.8c0,5.5-4.5,10.1-10.1,10.1c-5.6,0-10.1-4.6-10.1-10.1c0-5.6,4.5-10.1,10.1-10.1C84.2,46.7,88.7,51.3,88.7,56.8z};
  }
}

\newcommand\orcidicon[1]{\href{https://orcid.org/#1}{\mbox{\scalerel*{
\begin{tikzpicture}[yscale=-1,transform shape]
\pic{orcidlogo};
\end{tikzpicture}
}{|}}}}

\begin{document}

\title{Black hole in a generalized Chaplygin-Jacobi dark fluid: shadow and light deflection angle}

\author{Mohsen Fathi\orcidicon{0000-0002-1602-0722}}
\email{mohsen.fathi@ucentral.cl}
\affiliation{Vicerrector\'{i}a Acad\'{e}mica, Universidad Central de Chile,  Toesca 1783, Santiago 8320000, Chile}
\affiliation{Facultad de Ingenier\'{i}a y Arquitectura, Universidad Central de Chile, Av. Santa Isabel 1186, 8330601, Santiago, Chile}

\author{J.R.  Villanueva\orcidicon{0000-0002-6726-492X}}
    \email{jose.villanueva@uv.cl}
    \affiliation{Instituto de F\'{i}sica y Astronom\'{i}a, Universidad de Valpara\'{i}so,
Avenida Gran Breta\~{n}a 1111, Valpara\'{i}so, Chile}

\author{Gilberto Aguilar-P\'erez}
\email{gilaguilar@uv.mx}
    \affiliation{Facultad de F\'{\i}sica, Universidad Veracruzana 91097, Xalapa, Veracruz, M\'exico}

\author{Miguel Cruz\orcidicon{0000-0003-3826-1321}}
    \email{miguelcruz02@uv.mx}
    \affiliation{Facultad de F\'{\i}sica, Universidad Veracruzana 91097, Xalapa, Veracruz, M\'exico}

\begin{abstract}

We investigate a generalized Chaplygin-like gas with an anisotropic equation of state, characterizing a dark fluid within which a static spherically symmetric black hole is assumed. By solving the Einstein equations for this black hole spacetime, we explicitly derive the metric function. The spacetime is parametrized by two critical parameters, $\mB$ and $\alpha$, which measure the deviation from the Schwarzschild black hole and the extent of the dark fluid's anisotropy, respectively. We explore the behavior of light rays in the vicinity of the black hole by calculating its shadow and comparing our results with the Event Horizon Telescope observations. This comparison constrains the parameters to $0 \leq \mB \lesssim 0.03$ and $0 < \alpha \lesssim 0.1$. Additionally, we calculate the deflection angles to determine the extent to which light is bent by the black hole. These calculations are further utilized to formulate possible Einstein rings, estimating the angular radius of the rings to be approximately $37.6\,\mathrm{\mu as}$. Throughout this work, we present analytical solutions wherever feasible, and employ reliable approximations where necessary to provide comprehensive insights into the spacetime characteristics and their observable effects.

\bigskip

{\noindent{\textit{keywords}}: Dark energy, Chaplygin gas, black hole shadow, light deflection
}\\

\noindent{PACS numbers}: 04.20.Fy, 04.20.Jb, 04.25.-g   
\end{abstract}

\maketitle
\section{Introduction}

The dark side of the universe has profoundly influenced our physical observations, reshaping our understanding of phenomena such as flat galactic rotation curves, anti-lensing, and the accelerated expansion of the universe \cite{Rubin1980,Massey2010,Bolejko2013,Riess:1998,Perlmutter:1999,Astier:2012}. Since the late 20th century, two outstanding discoveries have advanced our comprehension of the cosmos. Firstly, the confirmation of highly isotropic black body radiation with temperature fluctuations on the order of $10^{-5}$, observed in the cosmic microwave background radiation (CMBR) \cite{Bennett_1994}. Secondly, the discovery of the universe's accelerated expansion, based on type Ia supernovae observations within the Friedmann-Lema\^{i}tre-Robertson-Walker (FLRW) metric \cite{Riess:1998,Perlmutter:1999}. These findings led to the development of the Lambda-Cold Dark Matter ($\Lambda$CDM) model. Despite its simplicity, the $\Lambda$CDM model accurately describes a vast array of observational data. However, its theoretical origins, particularly the interpretation and value of the cosmological constant, remain mysterious. Besides the current cosmological crisis well-known as $H_{0}$ tension, a central issue is the \textit{coincidence problem}: why does the cosmological constant's contribution match that of matter in our current epoch? Even extended models with dynamic sources have not provided a fundamental understanding of this component.

A promising approach to addressing the coincidence problem involves replacing the cosmological constant with a dynamical quintessence field, inspired by the inflaton field during the inflationary epoch. In modern cosmology, the universe's accelerated expansion poses a significant challenge, prompting the search for models to explain this phenomenon. Since dynamical fields appear to be the most promising scenario, various proposals have emerged as alternatives to the cosmological constant, including the Chaplygin Gas (CG) model \cite{Kamenshchik_2001}. The CG model, characterized by an exotic equation of state (EoS) $p=-A/\rho$, where $p$ is pressure and $\rho$ is energy density, offers a unique description of the transition from a matter-dominated universe to one experiencing accelerated expansion \cite{Kamenshchik_2001}. The stability of the CG model and its role in structure formation have been explored, highlighting its negative speed of sound and its initial phase resembling dust in the $\Lambda$CDM model \cite{Fabris_2002}. Extending the CG model, the Generalized Chaplygin Gas (GCG) introduces a parameter $\alpha$, resulting in an EoS $p=-{A}/{\rho^{\alpha}}$. This model maintains a connection to $d$-brane theories and evolves from non-relativistic matter to an asymptotically de Sitter phase \cite{Bento_2002}. Additionally, the GCG model aligns well with observations of inhomogeneities.

In studying astrophysical phenomena, adopting such models appears prudent given the lack of superior explanations. This includes phenomena like supernovae, galaxy clusters, quasars, and black holes. In particular, black holes have garnered significant interest, especially after the recent Event Horizon Telescope (EHT) imaging of M87* \cite{Akiyama:2019} and Sgr A* \cite{Akiyama:2022}, which demonstrated their observability beyond theoretical constructs. Cosmological dynamics also influence black hole evolution, particularly through dark components inferred from cosmological energy-momentum constituents. This includes considerations of dark matter halos \cite{pizzella_supermassive_2003,ferrarese_beyond_2002,sabra_black_2007,volonteri_how_2011,Xu:2018,konoplya_shadow_2019,jusufi_shadows_2020,xu_black_2020,hawkins_signature_2020,villanueva-domingo_brief_2021,Das:2021,liu_ringing_2021,liu_tidal_2022,pantig_dark_2022,pantig_dehnen_2022,pantig_black_2023,xavier_shadows_2023,vagnozzi_horizon-scale_2023,stelea_charged_2023,ovgun_constraints_2024,yang_black_2024,liu_modeling_2024,wu_rotating_2024,pantig_apparent_2024} and the coupling of spacetime with quintessential or alternative dark fields \cite{Kiselev:2003,JimenezMadrid:2005,maeda_self-similar_2008,Jamil:2009,mersini-houghton_investigating_2009,martin-moruno_dark_2009,cheng-yi_dark_2009,fernando_nariai_2013,babichev_black_2013,ghosh_rotating_2016,de_oliveira_three-dimensional_2018,fathi_study_2022,rogatko_dark_2024,singh_thermodynamic_2021,hui_echoes_2024}. These parameters must be calibrated based on standard observations.

Building on these discussions and alongside with the aforementioned interest, this work aims to study the optical properties of a static spherically symmetric black hole associated with a specific dark energy field, an extension of the GCG. To achieve this aim, we will first introduce this new theory and its EoS in Sect. \ref{sec:CJDF}. In Sect. \ref{sec:AdSBH_sol}, we will explore the density profile of this gas in a static spherically symmetric spacetime, rigorously examining its energy conditions, and solving the Einstein equations to derive the exterior geometry of a black hole immersed in the dark field. Furthermore, in Sect. \ref{sec:ShDe_gen}, we will discuss null geodesics in static black hole spacetimes, parametrizing the black hole shadow on the observer's screen and formulating the light deflection angle {using the general equation for planar orbits}. Section \ref{sec:SDcalc} will focus on measuring the shadow radius, illustrating shadow curves for different black hole parameters, calculating the light deflection angle, and examining its sensitivity to the anisotropy of the dark fluid. We will also compare our results with EHT observations to provide constraints on key black hole parameters, concluding with the angular radius of the Einstein ring. Finally, our findings will be summarized in Sect. \ref{sec:conclusions}. Throughout this paper, we will use geometric units with $8\pi G=c=1$ and the sign convention $(- + + \,+)$.

\section{Generalized Chaplygin-Jacobi Dark Fluid Overview}\label{sec:CJDF}

Considering the fundamental nature of the CG and its generalizations, they emerge as powerful elements in describing the dynamics of the universe across its various evolutionary stages \cite{Bili__2002}. However, when confronted with observational data, the original CG reveals inconsistencies that prompt the search for further generalizations. Beyond the GCG, the generalized Chaplygin-Jacobi gas (GCJG), proposed in Ref. \cite{Rengo1}, presents an alternative model in the context of inflationary cosmology. This model employs the Hamilton-Jacobi approach with the generating Hubble function
\begin{equation}\label{hcj}
    H(\bar{\phi}, \m) = H_0\, {\nc}^{\frac{1}{1+\alpha}}\Bigl([1+\alpha]\Phi \Bigr),
\end{equation}
where $\bar{\phi}$ is the scalar \textit{inflaton} field, $\Phi=\sqrt{6\pi/m_\mathrm{P}^2}\left(\bar{\phi}-\bar{\phi}_0\right)$ is a dimensionless quantity with $m_{\mathrm{P}}$ being the Planck mass, $H_0\equiv H(\bar{\phi}_0,\m)$, and  ${\nc}(x)= 1/{\cn}(x)$, with ${\cn}(x)
\equiv \cn(x; \m)$ being the Jacobi elliptic
cosine function with argument $x$ and modulus $\m$\footnote{The incomplete elliptic integral of the first kind is given by the general form  {\cite{handbookElliptic}}
\begin{eqnarray}\nonumber
    w={\bf{F}}(u,\m)=\int_0^u\frac{\ed t}{\sqrt{1-\m^2\sin^2{t}}},
\end{eqnarray}
where $u={\bf{F}}^{-1}(w,\m)={\bf{am}}(w,\m)$ is the Jacobi amplitude. This way, the Jacobi elliptic cosine function is defined as
\begin{eqnarray}\nonumber
\cos{u}=\cos{\big({\bf{am}}(w,\m)\big)=\cn(w,\m)}.
\end{eqnarray}}. 
In this theory, the EoS of the gas is obtained by means of the generating function \eqref{hcj} as follows \cite{Rengo1}:
\begin{equation}
\label{prro2}
p(\rho)=-\frac{B\,\m}{\rho^{\alpha}}-\m'\rho\left(2-\frac{1}{B}\rho^{\alpha+1}\right),
\end{equation}
where $B$ is a real constant, and $\m'=1-\m$ is the complementary modulus of the elliptic function. Note that for $\m=1$ (i.e., $\m'=0$), $\alpha=1$, and $B>0$, the gas becomes isotropic, and its EoS recovers $p=-B/\rho$, which is consistent with that on cosmological scales. Notably, in the limit $\m \rightarrow 1$, the GCG is recovered, and for $\m \neq 1$, the GCJG exhibits positive pressure. This interesting behavior persists even in the limit where $\m=0$, which corresponds to the trigonometric limit of the Jacobi elliptic functions. Therefore, the newly introduced parameter $\m$, along with other components of the theory, can enhance its utility in aligning with observational data and making the theory more reliable in explaining the accelerated expansion of the universe.

Beginning from the next section, we explore the properties of a spherically symmetric spacetime associated with the GCJG. We assume that the spacetime is filled with such a gas. Therefore, we prefer to regard this gas as a dark fluid. Henceforth, in this study, we will use the abbreviations CDF for the Chaplygin dark fluid and GCJDF for the generalized Chaplygin-Jacobi dark fluid.

\section{
(Anti-)de Sitter Black hole solution in the GCJDF background
}\label{sec:AdSBH_sol}

To explore the connection between black hole spacetimes and their surrounding fields, one needs to solve the gravitational field equations and the equations of motion for the relevant fields. Understanding the matter around black holes is challenging, in particular when it is influenced by exotic fields like quintessence dark energy and CG. Hence, exploring the relationship between matter and spacetime curvature using the fluid matter's EoS is crucial. For the present analysis, we adopt the static spherically symmetric spacetime metric
\begin{equation}
    \label{metr} 
    {\rm d} s^2=g_{\mu \nu} {\rm d}x^{\mu} {\rm d} x^{\nu}=-f(r) {\rm d} t^2+\frac{{\rm d} r^2}{g(r)}
    +r^2 \left( {\rm d}\theta^2 +\sin^2\theta {\rm d}\phi^2 \right),
\end{equation}
in the usual Schwarzschild coordinates $(t,r,\theta,\phi)$, where $f(r)$ and $g(r)$ are general radial-dependent functions. In cases where the surrounding matter can be regarded as a perfect fluid (i.e., the EoS remains $p=w \rho$ with $w=\text{const.}$), discussions have been done on black holes surrounded by dust, phantom energy, or dark energy \cite{Semiz:2011}. Furthermore, in the presence of a quintessential matter field, the fluid can be considered anisotropic, resulting in alterations to both the EoS and the relevant energy-momentum tensor from that of a perfect fluid \cite{kiselev_quintessence_2003}. Plausible options have been proposed for the background processing of the CDF, however, the main theoretical structure remains under debate \cite{nozari_observational_2011,benaoum_modified_2012,hulke_variable_2020,ray_anisotropic_2023}. Indeed, the CDF is modeled in terms of the scalar field $\varphi$ and the self-interacting potential $U(\varphi)$, contributing to the Lagrangian $\mathcal{L}_\varphi=-1/2\partial_a\varphi\partial^{a}\varphi-U(\varphi)$ \cite{debnath_role_2004,Mak:2005}. Since the GCJDF is also an anisotropic fluid, hence its energy-momentum tensor can be written as \cite{Raposo:2019} (see also Refs. \cite{Li:2023zfl,Arora:2023,Mustafa:2023})
\begin{equation} 
    {T_{\mu}}^{\nu} = \rho u_{\mu} u^{\nu}+p_r k_{\mu} k^{\nu}+ p_t {\Pi_{\mu}}^{\nu},
    \label{tmn}
\end{equation} 
where $p_r$ and $p_t$ represent, respectively, the radial and tangential components of the pressure, $\bm{u}$ is the four-velocity of the fluid, and $\bm{k}$ is a unit space-like vector orthogonal to $\bm{u}$. Hence, these vectors satisfy the conditions $\bm{u}\cdot\bm{u}=-1$, $\bm{k}\cdot\bm{k}=1$ and $\bm{u}\cdot\bm{k}=0$. Furthermore, the projection tensor onto a two-surface transverse to $\bm{u}$ and $\bm{k}$, is defined as ${\Pi_\mu}^{\nu}=\delta_\mu^\nu+u_\mu u^\nu-k_\mu k^\nu$, with $\delta_{\mu}^\nu$ being the Kronecker delta. In the comoving frame of the fluid, it is straightforward to get $u_{\alpha}= (-\sqrt{f}, 0, 0, 0)$ and ${k}_{\alpha}= (0, 1/\sqrt{g}, 0, 0)$. Accordingly, the energy-momentum tensor \eqref{tmn} can be re-expressed as
{\begin{equation}
     {T_\mu}^{\nu}=-(\rho+p_t)\delta_\mu^{t}
    \delta_{t}^{\nu}+p_t \delta_{\mu}^{\nu}+
    \Delta \,\delta_{\mu}^{r}
    \delta_{r}^{\nu},
    \label{tmn2}
\end{equation}
where $\Delta\equiv(p_r-p_t)$ indicates the difference between the components of the pressure and it is known as the anisotropic factor \cite{Li:2023zfl}. Hence, at the $p_r\rightarrow p_t$ limit, the energy-momentum tensor reduces to its standard isotropic form.} Note that, despite cosmological fluid anisotropy around a black hole, its EoS should still appear as $p=p(\rho)$ at cosmological scales, which allows for constraining the tangential pressure $p_t$ via isotropic averaging over the angles and employing some additional conditions such as $\langle {T_i}^j \rangle = p(\rho)\delta_i^j$. Therefore, one obtains the relation \cite{Li:2023zfl}
\begin{equation}\label{pptpr} 
    p(\rho)=p_t+\frac{1}{3}(p_r-p_t)=\frac{2 p_t}{3}+\frac{p_r}{3},
\end{equation}
by taking into account $\langle\delta_i^r\delta^j_r\rangle=1/3$. 
%
%
Note that, we require the extra condition $p_r=-\rho$ to conserve the staticity of the fluid and the continuity of its energy density across the black hole horizon. Accordingly, one can obtain the tangential pressure by using Eq. \eqref{pptpr}, which has been presented in Refs. \cite{kiselev_quintessence_2003, Li:2023zfl} for the cases of quintessential matter and the CDF. In the presence of the GCJDF, the above conditions together with the EoS in Eq. \eqref{prro2}, result in the relation 
\begin{equation}
    \label{prtang} 
    p_t=-\frac{3 B\,\m}{2 \rho^{\alpha}}-\frac{(6 \m'-1)}{2}\rho+
    \frac{3\m'}{2 B}\rho^{\alpha + 2},
\end{equation}
and hence, the components of the energy-momentum tensor are obtained as
\begin{eqnarray}
    && {T_{t}}^{t}={T_{r}}^{r}=p_r=-\rho,\label{radtens}\\
    && {T_{\theta}}^{\theta} =  {T_{\phi}}^{\phi}=p_t=-\frac{3 B \m}{2 \rho^{\alpha}}-\frac{(6 \m'-1)}{2}\rho+
    \frac{3\m'}{2 B}\rho^{\alpha + 2}. \label{angtens}
\end{eqnarray}
Since the condition ${T_{t}}^{t}={T_{r}}^{r}$ is desirable, one can assume $f(r)=g(r)$ in the line element \eqref{metr} without lose of generality. Accordingly, the components of the Einstein tensor are calculated as 
\begin{eqnarray}
   &&  {G_{t}}^{t} = G_{r}^{\,\,r}=\frac{f+r f'-1}{r^2},\label{radEtens}\\
   &&  {G_{\theta}}^{\theta} =  {G_{\phi}}^{\phi}=\frac{2 f'+r f''}{2 r},\label{angEtens}
\end{eqnarray}
in which $f'\equiv\partial_r f(r)$. Therefore, by utilizing Eqs. \eqref{radtens}--\eqref{angEtens}, the Einstein equations ${G_\mu}^\nu={T_\mu}^\nu$ yield the following differential equation system:
\begin{eqnarray}
    && \frac{f+r f'-1}{r^2} = -\rho,\label{eqein1}\\
    && \frac{2 f'+r f''}{2 r} =  -\frac{3 B \m}{2 \rho^{\alpha}}-\frac{(6 \m'-1)}{2}\rho+ \frac{3\m'}{2 B}\rho^{\alpha + 2}. \label{eqein2} 
\end{eqnarray}
When $B\equiv -\rho_0^{1+\alpha}\mathcal{B}<0$, where $\rho_0$ and $\mathcal{B}$ are positive constants, we have (see appendix \ref{app:A})
\begin{equation}
     \left(\frac{\rho}{\rho_0}\right)^{1+\alpha}=\mathcal{B} 
    \left[\frac{\m (1+y)-1}{\m' (1+y)}\right],
    \label{densid}
\end{equation}
in which $y\equiv y(r) = {q^2}/{r^{3(1+\alpha)}}$, with $q>0$ being a normalization factor that indicates
the intensity of the GCJDF. 
Note that, Eq. \eqref{densid} directly arises from the conservation law for the energy-momentum tensor, $\nabla_{\nu}{T_{\mu}}^\nu=0$. It can be verified that for small radial distances (i.e., $y\gg 1$), the energy density of the GCJDF is approximated by
\begin{equation}
\rho(r) \approx \rho_0 \mathcal{B}^{\frac{1}{1+\alpha}} \left[\frac{1-\m y  }{\m' y}\right]^{\frac{1}{1+\alpha}},
    \label{eq:rho_smallr}
\end{equation}
whereas for large distances (i.e., $y\ll 1$), it yields
\begin{equation}
\rho(r) \approx \rho_0\mathcal{B}^{\frac{1}{1+\alpha}} = (-B)^{\frac{1}{1+\alpha}}.
    \label{eq:rho_larger}
\end{equation}
The above result indicates that the GCJDF behaves similarly to a positive cosmological constant at large distances. Note that for $\alpha=1$, the result $\rho(r)\approx \sqrt{-B}$ is recovered, which aligns with that of the CDF on cosmological scales \cite{Li:2023zfl}. {In Table \ref{tab:1}, we calculate the large-distance limits of the pressure components for the GCJDF, showing that the anisotropic factor $\Delta = -\m' p_r$ does not tend to zero asymptotically. This behavior is different from that observed in the quintessential fluid \cite{kiselev_quintessence_2003} and the CDF \cite{Li:2023zfl}, where the anisotropy decreases with distance; for the GCJDF, the anisotropic factor only approaches zero when $\m$ tends to 1. This persistent anisotropy suggests various implications for the universe's evolution, such as unaccounted X-ray absorption, bulk flows, and Hubble expansion anisotropies. Observational studies, like those based on galaxy clusters \cite{M18, M20, M21} and type Ia supernovae (SN Ia) in the context of Lema\^{i}tre–Tolman-Bondi (LTB) models \cite{DelCampo:2012wkb}, support these ideas. These findings open up opportunities for further investigation of the GCJDF, guided by the referenced studies and others \cite{Amirhashchi:2018bic, Anisotp1, Devi:2022pgi, ray_anisotropic_2023}.}
\begin{table}[t]
    \centering
    \begin{tabular}{@{}c||c|c|c|c|c@{}}
         anisotropic fluid  &  EoS  & $p_r$ & $p_t$ & $\rho$ & asymptotic behavior \\
         \hline\hline
        quintessential matter \cite{kiselev_quintessence_2003} & $p=w \rho,\,\, (-1<w<-1/3)$  & $-\rho$ & $\frac{1}{2}(1+3w)\rho$ & $-\frac{a}{2}\frac{3w}{r^{3(w+1)}}$ & $\begin{array}{lcl}
        \rho&\rightarrow&0\\
        p_r&\rightarrow&0\\
        p_t&\rightarrow&0
        \end{array}$ \\
        \hline
         CDF \cite{Li:2023zfl} & $p=-{B}/{\rho},\,\,(B>0)$ & $-\rho$ & $\frac{\rho}{2}-\frac{3B}{2\rho}$ & $\sqrt{B+\frac{q^2}{r^6}}$ & $\begin{array}{lcl}
         \rho&\rightarrow&\sqrt{B}\\
         p_r&\rightarrow&-\sqrt{B}\\ 
         p_t&\rightarrow&-\sqrt{B}
         \end{array}$ \\
         \hline
         GCJDF & $\begin{array}{lcl}
         p&=&-\frac{B\,\m}{\rho^{\alpha}}\\&&-\m'\rho\left(2-\frac{1}{B}\rho^{\alpha+1}\right),\,\, (B<0)
         \end{array}$ & $-\rho$  & $\begin{array} {lcl}
         -\frac{3 B \m}{2 \rho^{\alpha}}-\frac{(6 \m'-1)}{2}\rho\\+\frac{3\m'}{2 B}\rho^{\alpha +2}
         \end{array}$ &  $  
    \left[\frac{B(\m'-\m y)}{\m' (1+y)}\right]^{\frac{1}{1+\alpha}}$  & $\begin{array}{lcl}
    \rho&\rightarrow& (-B)^{\frac{1}{1+\alpha}}\\
    p_r&\rightarrow&-(-B)^{\frac{1}{1+\alpha}}\\
    p_t&\rightarrow& {(2-\m) p_r}
    \end{array}$\\
    \end{tabular}
    \caption{Characteristics of quintessential matter, CDF, and the GCJDF.}
    \label{tab:1}
\end{table}

\begin{figure}[t]
    \centering
    \includegraphics[width=7cm]{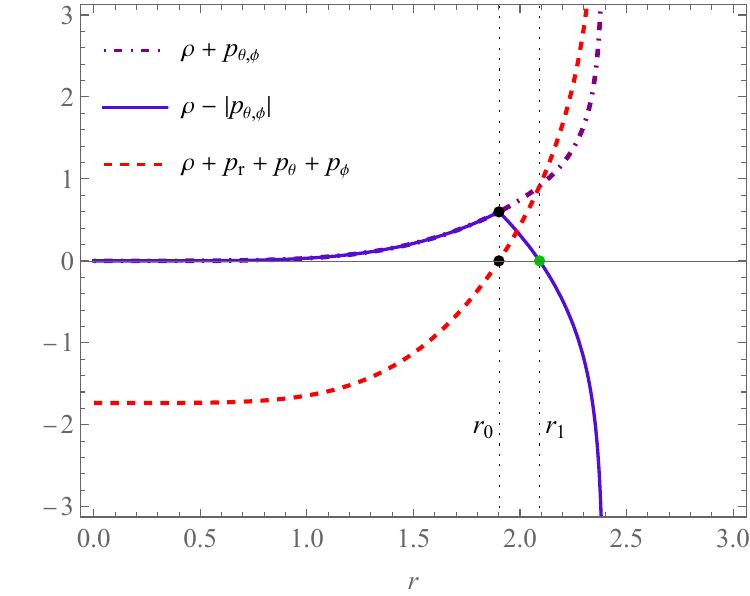}
    \caption{The radial profiles of the values in Eqs. \eqref{eq:EConds1} are plotted for $\mathcal{B}=0.03$, $\alpha=0.6$, $\m=0.4$, $q=10$, and $\rho_0=10$. The quantity $\rho+p_{\theta,\phi}$ remains positive throughout the domain, while $\rho+p_r+p_\theta+p_\phi$ shifts from negative to positive at $r_0=1.901$, denoted by black points. This transition point corresponds to the sign change of $p_{\theta,\phi}$ from attractive to repulsive, where $\rho-|p_{\theta,\phi}|$ shifts from ascending to descending, until it finally enters the domain of negative values at $r_1=2.092$ (the green point).
      }
    \label{fig:EConds}
\end{figure}

Examining the energy conditions for the GCJDF is also crucial. Energy conditions serve as essential tools for scrutinizing cosmological geometry \cite{visser_energy_2000} and black hole spacetimes \cite{zaslavskii_regular_2010,neves_regular_2014} within both general relativity \cite{kontou_energy_2020} and modified gravity \cite{capozziello_energy_2014}. These conditions include the null energy condition (NEC), weak energy condition (WEC), strong energy condition (SEC), and dominant energy condition (DEC), each defined as \cite{Li:2023zfl}
\begin{itemize}

    \item \underline{NEC}: $\rho+p_i\geq 0$,

    \item \underline{WEC}: $\rho\geq0$ \quad and \quad $\rho+p_i\geq0$,

    \item \underline{SEC}: $\rho+\sum_i p_i\geq0$ \quad and \quad $\rho + p_i \geq 0$,

    \item \underline{DEC}: $\rho \geq0$\quad and \quad $|p_i|\leq\rho$,
    
\end{itemize}
in which $i=\{r,\theta,\phi\}$. Applying Eqs. \eqref{prtang}--\eqref{angtens} for the GCJDF, the relevant values are calculated as 
\begin{subequations}
    \begin{align}
        & \rho+p_r=0,\qquad \rho+p_{\theta,\phi}=\frac{3}{2} \left(\rho-B \m \rho ^{-\alpha }+\frac{\m' \rho ^{\alpha +2}}{B}-2 \m' \rho  \right),\\
        & \rho+p_r+p_\theta+p_\phi=\rho-3 B \m \rho ^{-\alpha }+\frac{3 \m' \rho ^{\alpha +2}}{B}-6 \m' \rho ,\\
        & \rho-|p_r|=0,\qquad \rho-|p_{\theta,\phi}|=\rho -\frac{1}{2} \left|\rho-3 B \m \rho ^{-\alpha }+\frac{3 \m' \rho ^{\alpha +2}}{B}-6 \m' \rho \right|.
    \end{align}
    \label{eq:EConds1}
\end{subequations}
In Fig. \ref{fig:EConds}, we have graphed the radial profiles of unspecified parameters described in Eqs. \eqref{eq:EConds1}, by using Eq. \eqref{densid}, to analyze the energy conditions of the GCJDF in our scenario.  
Based on the behavior of these quantities, one can infer the following:
\begin{subequations}
    \begin{align}
    & \rho + p_{\theta,\phi} \geq 0 \quad\Longrightarrow\quad \text{NEC and WEC are satisfied},\\
    & \rho+\sum_{i=r,\theta,\phi}p_i\left\{
        \begin{array}{cc}
             \leq 0\quad\text{for}\quad r\leq r_0\quad\Longrightarrow\quad \text{SEC is violated,} \\
             > 0\quad\text{for}\quad r> r_0\quad\Longrightarrow\quad \text{SEC is satisfied,}
        \end{array}
        \right.\\
    & \rho-|p_{\theta,\phi}|\left\{
        \begin{array}{cc}
             \geq 0\quad\text{for}\quad r\leq r_1 \quad\Longrightarrow\quad \text{DEC is satisfied,} \\
             < 0\quad\text{for}\quad r> r_1 \quad\Longrightarrow\quad \text{DEC is violated.}
        \end{array}
        \right.
    \end{align}
\end{subequations}
Note that, having both repulsive gravitational effects and obeying the SEC is not inherently contradictory in general relativity. The SEC mainly ensures that gravity overall pulls things together. However, when the DEC is violated, it means certain types of energy, like dark energy, become dominant, leading to repulsive gravitational effects, especially at large distances. This fits with what we observe in the universe's expansion. For instance, in some cosmological models with exotic matter like phantom energy, we see both SEC compliance and DEC violation, causing the universe's expansion to speed up in unexpected ways.
As declared in Ref. \cite{BARROW1988743}, the SEC is obeyed by the matter source in deflationary models, but it is violated during the early evolution of the cosmos in such models, allowing for the avoidance of the initial singularity. Conversely, violation of the DEC corresponds to the violation of the generalized second law of thermodynamics in general relativity. However, such violation does not necessarily invalidate a given cosmological model (see, e.g., Ref. \cite{2008PhDT}). Similarly, violation of the SEC is not in contrast with ordinary inflationary models \cite{BARROW198712}. In general, violation of the DEC is attributed to the bulk viscous stress as a result of particle production \cite{BARROW1988743,giovannini_relic_1999,caldwell_phantom_2002,lasukov_violation_2020}. However, situations where the SEC is satisfied while the DEC is violated are not difficult to find. An example would be during the transition from the Schwarzschild vacuum to de Sitter spacetime, as discussed in Ref. \cite{conboy_smooth_2005}. Based on the above notes, it is reasonable to anticipate the presence of an exotic field outside the Schwarzschild radius at $r_s=2$ in our model.

\begin{figure}[t]
    \centering
    \includegraphics[width=8cm]{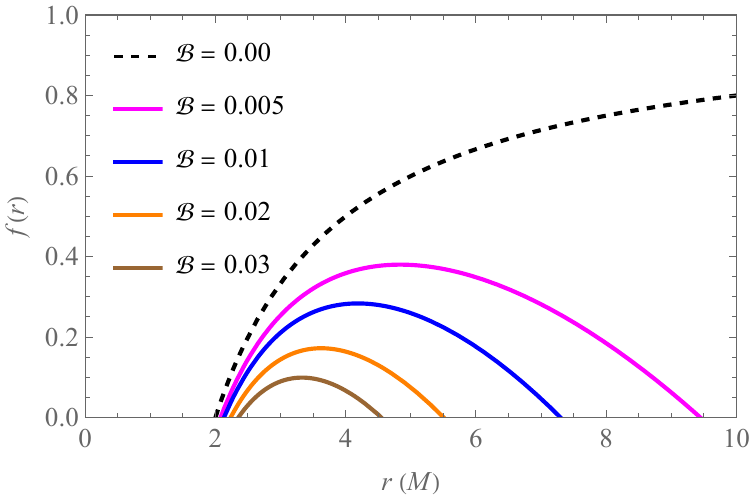} 
    \caption{The radial profile of the lapse function of the GCJDF black hole plotted for $\alpha=0.6$, $q=0.1$, $\m=0.4$, $\rho_0=1$, and various values of the $\mathcal{B}$-parameter. The dashed curve corresponds to the case of the Schwarzschild black hole with $\mathcal{B}=0$. In this diagram and all subsequent ones within this paper, the unit of length along the axes is considered as the black hole mass $M$.}
    \label{fig:f(r)}
\end{figure}

\begin{figure}[t]
    \centering
    \includegraphics[width=8cm]{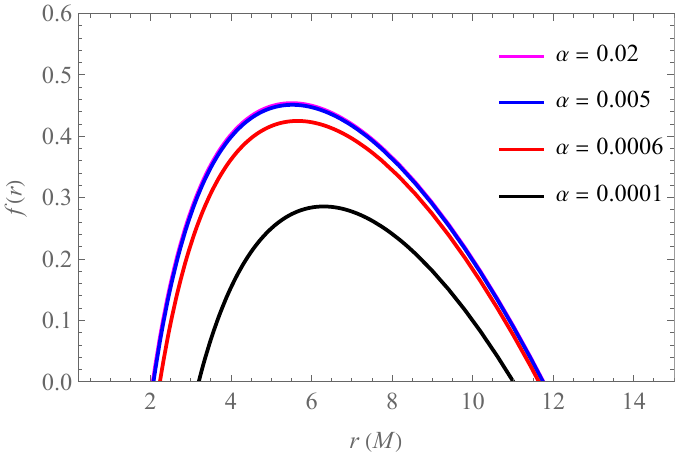}
    \caption{The radial profile of the reduced lapse function with the cosmological parameter given in Eq. \eqref{eq:lambda_red}, plotted for $\mathcal{B}=0.03$, $q=0.1$, $\m=0.4$, $\rho_0=1$, and some small values of the $\alpha$-parameter.}
    \label{fig:f(r)_red}
\end{figure}

Now, substitution of the expression \eqref{densid} in the differential equation \eqref{eqein1} results in the (anti-)de Sitter-like (Ads-like) lapse function
\begin{equation}\label{lapf01} 
    f(r)=1-\frac{2 M}{r}-\frac{1}{3}
    \lambda(r) r^2,
    \end{equation}
where the function $\lambda(r)$ is
\begin{equation}
    \label{lambfunc}
    \lambda(r)=\rho_0 \left[\frac{\mathcal{B}\bigr(1+y(r)\bigr)\bigr(\m y(r)-\m'\bigr)}{1+\frac{\m}{\m'}y(r)}\right]^{\frac{1}{1+\alpha}}
    \times F_1\left(-\frac{1}{1+\alpha};-\frac{1}{1+\alpha},\frac{1}{1+\alpha};\frac{\alpha }{1+\alpha };\frac{\m }{\m'} y(r),-y(r)\right),
\end{equation}
in which, $F_1(a; b_1, b_2; z; x_1, x_2)$ represents the two-variable Appell hypergeometric function. The integration constant in Eq. \eqref{lapf01} is set to $2M$ to retrieve the Schwarzschild solution component (by setting $\rho_0=0$ or $\mathcal{B}=0$), where $M$ denotes the black hole mass. It is important to note that we cannot recover the CDF solution at the limit of $\alpha=1=\m$, as this solution is well-defined only for $\m'\neq0$. To retrieve that solution, one would need to start from the corresponding EoS and the density profile.
In fact, the Appell function in Eq. \eqref{lambfunc} can be transformed as \cite{schneider_multiple_2013}
\begin{multline}
F_1\left(-\frac{1}{1+\alpha};-\frac{1}{1+\alpha},\frac{1}{1+\alpha};\frac{\alpha }{1+\alpha };\frac{\m }{\m'} y(r),-y(r)\right) \\
=\left[1-\frac{\m}{\m'}y(r)\right]^{\frac{1}{1+\alpha}}F_1\left(-\frac{1}{1+\alpha};\frac{\alpha}{1+\alpha},\frac{1}{1+\alpha};\frac{\alpha }{1+\alpha };\frac{y(r)}{\m y(r)-\m'},\frac{y(r)}{\m y(r)-\m'}\right)\\
= \left[1-\frac{\m}{\m'}y(r)\right]^{\frac{1}{1+\alpha}}
\, _2F_1\left(1,-\frac{1}{\alpha +1};\frac{\alpha }{\alpha +1};\frac{y(r)}{\m y(r)-\m'}\right),
    \label{eq:F1trans}
\end{multline}
in which $_2F_1(a,b;z,x)$ is the Gaussian hypergeometric function. Based on the inclusion of the simpler hypergeometric function in this relation, the computational time for numeric processes is significantly reduced. Consequently, the form presented in Eq. \eqref{eq:F1trans} will be considered for further studies.
In Fig. \ref{fig:f(r)}, the radial profile of the lapse function \eqref{lapf01} has been plotted for some values of the parameter $\mathcal{B}$. As expected, the Schwarzschild black hole with $\mathcal{B}=0$ shows the largest values for the lapse function. On the other hand, the GCJDF black hole possesses two horizons; the event horizon $r_+$ (where the infinite redshift occurs), and the cosmological horizon $r_{++}$ (where the infinite blueshift occurs). Accordingly, the lapse function has an extremum and its value decreases by the increase in the $\mathcal{B}$-parameter.
Before proceeding with further studies on the GCJDF black hole, it is worth mentioning that for small values of $\alpha$, where $\alpha+1\approx 1$, the hypergeometric function in Eq. \eqref{lambfunc} can be simplified to
\begin{equation}
    F_1\left(-\frac{1}{1+\alpha};-\frac{1}{1+\alpha},\frac{1}{1+\alpha};\frac{\alpha }{1+\alpha };\frac{\m }{\m'} y(r),-y(r)\right)\approx 1+\frac{q^2 }{\alpha\m'r^{3}},
    \label{eq:F1reduced}
\end{equation}
by taking into account the approximation $y(r)\approx q^2 r^{-3}$ for $\alpha\ll 1$. In this limit, therefore, the parameter $\lambda(r)$ in Eq. \eqref{lambfunc} is approximated as
\begin{equation}
\lambda(r) \approx \frac{\text{$\rho_0$} \mathcal{B} \left(q^2+r^3\right) \left(\m q^2-\m' r^3\right) \left(q^2+\alpha  \m' r^3\right)}{\alpha  r^6 \left(\m' r^3+\m q^2\right)} +\mathcal{O}(\alpha^2).
    \label{eq:lambda_red}
\end{equation}
In Fig. \ref{fig:f(r)_red}, the radial profile of the lapse function $f(r)$ has been plotted, by considering the cosmological parameter in Eq. \eqref{eq:lambda_red}.
One can observe from the profiles that, for $\alpha>0.001$, the changes in the maximum of $f(r)$ is not great. This domain is also more reliable for the condition $\alpha+1\approx 1$, for which, $\alpha$ is still have values notably larger than zero. We can therefore infer that for this limit, the alternations in the black hole spacetime, receive more contributions from changes in the $\mathcal{B}$-parameter. 

\begin{figure}[t]
    \centering
    \includegraphics[width=8cm]{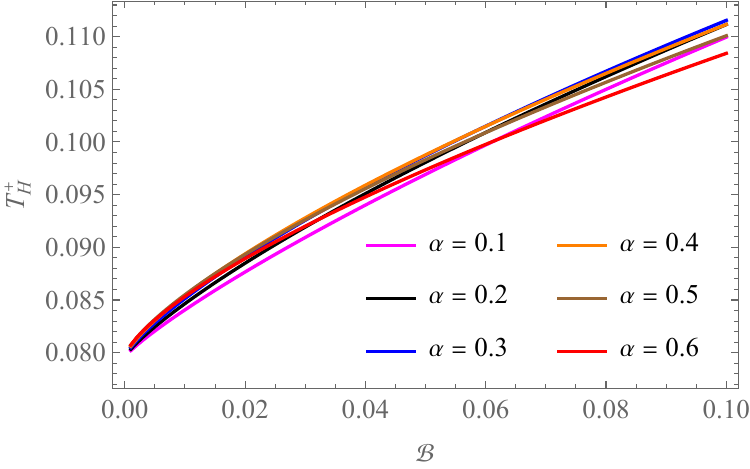}
    \caption{The $\mB$-profile of the Hawking temperature on the event horizon, $T^+_\mathrm{H}$, for different values of the $\alpha$-parameter, considering $q=0.1$, $\m=0.4$ and $\rho_0=1$.}
    \label{fig:TH}
\end{figure}

It is also important to note that besides the natural curvature singularity at $r=0$, there is another singularity occurring at
\begin{equation}
r_* = \left(\frac{\m q^2}{\m'}\right)^{\frac{1}{3(\alpha+1)}},
    \label{eq:rstar}
\end{equation}
at which $f(r)$ diverges. It is, in general, required that the curvature singularity be located inside the event horizon, in other words, $r_*<r_+$, which is respected by the GCJDF black hole. Let us now select a Bondi coordinate $v$ such that it aligns with the center of mass Bondi cuts in the limit $v\rightarrow\infty$. This choice pertains to the past of an open set of future null infinity $\mathcal{J}^+$, defined by points whose Bondi retarded time is defined in the domain $v\in(v_0,\infty)$ \cite{Gallo:2021}. The null geodesic congruence, defined by $\tilde\ell=\ed v$ permits the introduction of the affine parameter $r$, which in our black hole spacetime serves as the radial coordinate. This parameter is chosen such that it asymptotically matches the luminosity distance. The surfaces $(v,r)=\mathrm{const.}$ represent spheres, inheriting natural spherical coordinates from the Bondi cuts at $\mathcal{J}^+$, which label null rays of the congruence. Collectively, these elements establish a coordinate system $(v,r,x^{A})$, with $x^{A}$ denoting coordinates on the two-dimensional sphere $S^2$ \cite{Gallo:2021}. This way, the line element of the GCJDF black hole can be recast as
\begin{equation}
\ed s^2=-f(r)\ed v+2\ed v \ed r+r^2\left(\ed\theta^2+\sin^2\theta\ed\phi^2\right),
    \label{eq:metr_bondi}
\end{equation}
in the Bondi coordinates, possessing the time-like Killing vector $\xi^\mu=(1,0,0,0)$. Accordingly, if the above line element is applied, we obtain $\xi_\mu=\big(f(r),1,0,0\big)$. Now to obtain the surface gravity of the hypersphere corresponding to the black hole event horizon, we re-call the relation ${\xi^\mu}_{;\nu}\xi^{\nu}=\kappa \xi^\mu$ as the non-affinely parametrized geodesics equations, in which $\kappa$ is the surface gravity \cite{nielsen_dynamical_2008,Pielahn:2011}. Applying the metric \eqref{eq:metr_bondi}, this equation results in 
\begin{equation}
\kappa_+=\frac{1}{2}f'(r_+),
    \label{eq:kappa+}
\end{equation}
as the surface gravity of the black hole on its event horizon. Hence, the corresponding Hawking temperature of the GCJDF is obtained as $T_{\mathrm{H}}^+=\kappa_+/(2\pi)$. In Fig. \ref{fig:TH}, the $\mB$-profile of the Hawking temperature of the GCJDF black hole has been demonstrated, for some different values of the $\alpha$-parameter. This diagram has been obtained by generating the triplet $\{\mB,r_+,T_{\mathrm{H}}^+\}$, for  fixed values for the other parameters. 
The plot shows the variation of $T_{\mathrm{H}}^+$ with respect to $\mB$ for different values of $\alpha$. Starting from the Schwarzschild black hole (i.e. $\mB=0$), $T_{\mathrm{H}}^+$ rises with increasing $\mB$. It is evident that as $\alpha$ rises, the rate of increase in $T_{\mathrm{H}}^+$ with respect to $\mB$ also escalates.

Now that the general structure of the black hole in the GCJDF has been identified, we commence our investigation on the properties of this black hole from the next section.

\section{General formalism for shadow formation and light deflection angle}\label{sec:ShDe_gen}

In the presence of a black hole, light emitted from a source may undergo deflection due to the gravitational field of the black hole before reaching an observer. This deflection can lead to the formation of a shadow, where some photons are absorbed by the black hole, creating a region devoid of light. The boundary of this shadow defines the apparent shape of the black hole. Here, we review the general equations required to calculate the shape of the shadow, the energy emission rate, and the deflection angle. These formulas are derived for the general ansatz \eqref{metr} for the case of $f(r)=g(r)$, necessitating an examination of test particle motion within the spacetime.

\subsection{The equations of motion for light rays}\label{subsec:SheDe_gen_1}

Pursuing the standard Lagrangian method in the study of test particle motion in gravitational fields, let us write the Lagrangian \cite{Chandrasekhar:579245}
\begin{equation}
    \mathcal{L} = \frac{1}{2}g_{\mu\nu} \dot{x}^\mu \dot{x}^\mu,
    \label{eq:L}
\end{equation}
in which the overdot represents differentiation with respect to the affine parameter of the geodesic curves, denoted as $\tau$. This way, the components of the canonically conjugate momentum can be defined as 
\begin{eqnarray}
    && p_t = f(r) \dot{t} = E,\label{eq:pt}\\
    && p_r = \frac{\dot{r}}{f(r)},\label{eq:pr}\\
    && p_\theta = r^2 \dot{\theta},\label{eq:ptheta}\\
    && p_\phi = r^2 \sin^2\theta \dot{\phi} = L,\label{eq:pphi}
\end{eqnarray}
in which $E$ and $L$ are, respectively, the energy and the angular momentum associated to the test particles. This way, the Hamilton-Jacobi equation can be written as \cite{Carter:1968}
\begin{equation}
\frac{\partial\mathcal{S}}{\partial\tau}=-\frac{1}{2}g^{\mu\nu}\frac{\partial\mathcal{S}}{\partial x^\mu}\frac{\partial\mathcal{S}}{\partial x^\nu},
    \label{eq:H-J0}
\end{equation}
where $\mathcal{S}$ is the Jacobi action. Taking into account the ansatz \eqref{metr} in the above equation, we can recast the Hamilton-Jacobi equation as
\begin{equation}
-2\frac{\partial\mathcal{S}}{\partial\tau} = -\frac{1}{f(r)}\left(\frac{\partial\mathcal{S}_t}{\partial t}\right)^2 + {f(r)}\left(\frac{\partial\mathcal{S}_r}{\partial r}\right)^2
+\frac{1}{r^2}\left(\frac{\partial\mathcal{S}_\theta}{\partial \theta}\right)^2
+\frac{1}{r^2 \sin^2\theta}\left(\frac{\partial\mathcal{S}_\phi}{\partial \phi}\right)^2.
    \label{eq:H-J1}
\end{equation}
Based on the method of the separation of variables, and considering the constants of motion defined earlier, the Jacobi action can be expressed as
\begin{equation}
\mathcal{S}= \frac{1}{2}m^2\tau - E t+L\phi + \mathcal{S}_r(r)+\mathcal{S}_\theta(\theta),
    \label{eq:S}
\end{equation}
with $m$ being the rest mass of the test particles. Since in this study we are dealing with photons, we hence set $m=0$. Now using Eq. \eqref{eq:S} in Eq. \eqref{eq:H-J1}, we get
\begin{equation}
0 = \frac{E^2}{f(r)}-f(r)\left(\frac{\partial\mathcal{S}_r}{\partial r}\right)^2-\frac{1}{r^2}\left[\frac{L^2}{\sin^2\theta}+\Q-L^2\cot^2\theta\right]
-\frac{1}{r^2}\left[\left(\frac{\partial\mathcal{S}_\theta}{\partial\theta}\right)^2-\Q+L^2\cot^2\theta\right],
    \label{eq:H-J2}
\end{equation}
where $\Q$ is the Carter's constant. This way, the following two equations are obtained:
\begin{eqnarray}
    && r^4 f(r)^2\left(\frac{\partial\mathcal{S}_r}{\partial r}\right)^2=r^4E^2-r^2\left(L^2+\Q\right)f(r),\label{eq:H-J3a}\\
    && \left(\frac{\partial\mathcal{S}_\theta}{\partial \theta}\right)^2 = \Q - L^2\cot^2\theta.\label{eq:H-J3b}
\end{eqnarray}
By applying Eq. \eqref{eq:pt}--\eqref{eq:pphi} to the above equations,  we get to the full set of equations of motion for null geodesics as
\begin{eqnarray}
    && \dot t = \frac{E}{f(r)},\label{eq:tdot}\\
    && \dot r =\pm\frac{\sqrt{\R(r)}}{r^2},\label{eq:rdot}\\
    && \dot\theta = \pm\frac{\sqrt{\Theta(\theta)}}{r^2},\label{eq:thetadot}\\
    && \dot\phi = \frac{L}{r^2\sin^2\theta},
    \label{eq:phidot}
\end{eqnarray}
in which the $+\, (-)$ sign corresponds to the outgoing (ingoing) nature of the trajectories. In the above relations, we have defined the variables
\begin{subequations}
    \begin{align}
        & \R(r) = r^4E^2-r^2\left(L^2+\Q\right)f(r),\label{eq:R}\\
        & \Theta(\theta) = \Q - L^2\cot^2\theta.\label{eq:Theta}
    \end{align}
\end{subequations}
Note that, by considering the definition \eqref{eq:R}, we can recast Eq. \eqref{eq:rdot} as
\begin{equation}
\dot r^2 + V(r) = 0,
    \label{eq:rdot_1}
\end{equation}
in which 
\begin{equation}
V(r) = \frac{f(r)}{r^2}\left(L^2+\Q\right)-E^2,
    \label{eq:V}
\end{equation}
is the effective potential for the null geodesics. Indeed, photons approaching the black hole may either deflect from it or plunge onto the event horizon. These potential outcomes for photons can be predicted by considering their initial energy and angular momentum in the gravitational effective potential. Furthermore, the point at which photons reach the threshold between these two fates corresponds to the maximum of this effective potential, which can be determined based on specific criteria 
\begin{equation}
    V(r_p)=V'(r_p)=0,\qquad \R(r_p)=\R'(r_p)=0.\label{eq:Vp}
\end{equation}
Here, $r_p$ denotes the radius of the photon sphere around the black hole, where photons reside in a state of instability. Therefore, one can deduce that the radius of the photon sphere can be determined as the smallest root of the equation
\begin{equation}
r_p f'(r_p) - 2 f(r_p)=0.
    \label{eq:rpEq}
\end{equation}

\subsection{Shadow parametrization}\label{subsec:shadowparam}

The test particle geodesics can be identified by the two impact parameters \cite{Bardeen:1973b,Chandrasekhar:579245}
\begin{eqnarray}
    && \xi = \frac{L}{E},\label{eq:xi}\\
    && \eta = \frac{\Q}{E^2}.\label{eq:eta}
\end{eqnarray}
Accordingly, one can recast
\begin{eqnarray}
    && V(r) = E^2\left[\frac{f(r)}{r^2}\left(\eta+\xi^2\right)-1\right],\label{eq:V_1}\\
    && \R(r)=E^2 r^2\left[r^2-f(r)\left(\eta+\xi^2\right)\right].\label{eq:R_1}
\end{eqnarray}
Hence, the conditions \eqref{eq:Vp} provide the expressions
\begin{eqnarray}
    && \xi_p^2(r_p) = \frac{r_p^2\left[2f(r_p)-r_pf'(r_p)\right]}{2f(r_p)\left[2f(r_p)+r_p f'(r_p)\right]},\label{eq:xip2}\\
    && \eta_p(r_p) = \frac{r_p^2\left[6 f(r_p)+r_p f'(r_p)\right]}{2f(r_p)\left[2f(r_p)+r_p f'(r_p)\right]},\label{eq:etap}
\end{eqnarray}
that satisfy the relationship
\begin{equation}
\eta_p+\xi_p^2 = \frac{4 r_p^2}{2f(r_p)+r_p f'(r_p)},
    \label{eq:etap+xip2}
\end{equation}
on the photon sphere. As mentioned earlier, the photon sphere is formed by photons on unstable orbits. In fact, photons orbiting on the unstable trajectories within the gravitational field of black holes will either plunge into the event horizon or escape from it. Those that escape, form a luminous photon ring that encapsulates the shadow of the black hole \cite{Synge:1966,Cunningham:1972,Bardeen:1973a,Luminet:1979}. Notably, Luminet's optical simulation of a Schwarzschild black hole and its accretion in 1979 \cite{Luminet:1979} provided deeper insights into these photon rings, originating from the highly distorted region around black holes. The resulting formulations aided scientists in confining the shadow of rotating black holes within their respective photon rings. Subsequently, Bardeen developed mathematical techniques to compute the shape and size of a Kerr black hole's shadow \cite{Bardeen:1972a,Bardeen:1973a,Bardeen:1973b}, later refined and applied by Chandrasekhar \cite{Chandrasekhar:579245}. These methods were further expanded and generalized extensively (see Refs. \cite{Bray:1986,Vazquez:2004,Grenzebach:2014,Grenzebach:2016,Perlick:2018a,Kogan:2018}). With these tools, rigorous analytical calculations, simulations, numerical and observational studies were conducted on a multitude of black hole spacetimes, including those with cosmological components \cite{Vries:1999,Shen:2005,Amarilla:2010,Amarilla:2012,Yumoto:2012,Amarilla:2013,Atamurotov:2013,Abdujabbarov:2015,Abdujabbarov:2016,Amir:2018,Tsukamoto:2018,Cunha:2018,Mizuno:2018,Mishra:2019,Kumar:2020b}. The black hole shadow is of significant importance as it offers insights into light propagation in the vicinity of the horizon. Recent investigations have aimed to establish relationships between the shadow and black hole parameters (see, e.g., Refs. \cite{Zhang:2020,Belhaj:2020} for thermodynamic aspects and Refs. \cite{Kramer:2004,Psaltis:2008,Harko:2009,Psaltis:2015,Johannsen:2016a,Psaltis:2019,Dymnikova:2019,Kumar:2020a} for dynamical aspects of black hole shadows). To determine the shadow of the GCJDF black hole, we must consider that this spacetime is not asymptotically flat. Hence, the traditional approach of considering an observer at infinity, as done in Refs. \cite{Bardeen:1973a,Bardeen:1973b} and Ref.~\cite{Vazquez:2004}, cannot be applied. Instead, an observer is located at the coordinate position $(r_o,\theta_o)$, characterized by the orthonormal tetrad $\bm{e}_{\{A\}}$, expressed as
\begin{subequations}
    \begin{align}
        & \bm{e}_0 = \left.\frac{1}{\sqrt{f(r)}}\,\partial_t\right|_{(r_o,\theta_o)},\label{eq:e0}\\
        & \bm{e}_1 = \left.\frac{1}{r}\,\partial_\theta\right|_{(r_o,\theta_o)},\label{eq:e1}\\
        & \bm{e}_2 = \left.-\frac{\csc\theta}{r}\,\partial_\phi\right|_{(r_o,\theta_o)},\label{eq:e2}\\
        & \bm{e}_3 = \left.-\sqrt{f(r)}\,\partial_r\right|_{(r_o,\theta_o)},\label{eq:e0}
    \end{align}
    \label{eq:eA}
\end{subequations}
which satisfy the condition ${{e}_A}^\mu {{e}^B}_\mu = \delta^B_A$. 
The method, initially proposed in Ref. \cite{Grenzebach:2014}, facilitates the computation of celestial coordinates in spacetimes featuring cosmological constituents (also discussed in Ref. \cite{Grenzebach:2016}). Within the aforementioned set of tetrads, the time-like vector $\bm{e}_0$ serves as the velocity four-vector of the chosen observer. Additionally, $\bm{e}_3$ is oriented towards the black hole, and $\bm{e}_0\pm\bm{e}_3$ generates the principal null congruence. Consequently, a linear combination of $\bm{e}_{\{A\}}$ aligns with the tangent to the light ray $\bm\ell(\tau) = \left(t(\tau),r(\tau),\theta(\tau),\phi(\tau)\right)$, originating from the black hole and extending into the past. This tangent can be parametrized in the two distinct ways
\begin{eqnarray}\label{eq:elldot}
 &&  \dot{\bm{\ell}} ={\dot t}{} \, \partial_t + {\dot r}\,\partial_r+{\dot\theta}\, \partial_\theta + {\dot\phi}\, \partial_\phi,\\
 &&   \dot{\bm{\ell}} = -\frac{E}{\sqrt{f(r)}}\bigl(-\bm{e}_0+\sin\vartheta\cos\psi\,\bm{e}_1 + \sin\vartheta\sin\psi\,\bm{e}_2 + \cos\vartheta\,\bm{e}_3\bigr),
\end{eqnarray}
in which $\vartheta$ and $\psi$ represent newly defined celestial coordinates in the observer's sky. It is evident that $\vartheta = 0$ directly aligns with the black hole's axis. As the boundary curve of the shadow arises from light rays intersecting the unstable (critical) null geodesics at the radial distance $r_p$, this region corresponds to the critical impact parameters $\xi_p$ and $\eta_p$, as given in Eqs.~\eqref{eq:xip2} and \eqref{eq:etap}. The respective celestial coordinates $(\psi_p,\vartheta_p)$ at this distance, for an observer positioned at $(r_o,\theta_o)$, have been derived in Ref. \cite{Grenzebach:2014} as
\begin{eqnarray}\label{eq:vartheta,psi}
    && \mathcal{P}(r_p,\theta_o) \vcentcolon= \sin\psi_p = \frac{\csc\theta_o \,\xi_p(r_p)}{\sqrt{\eta_p(r_p)}},\label{eq:psi_p}\\
     &&  \mathcal{T}(r_p,r_o) \vcentcolon= \sin\vartheta_p = \frac{1}{r_o} {\sqrt{ f(r_o)\, \eta_p(r_p)}}.\label{eq:vartheta_p}
\end{eqnarray}
Hence, the $\vartheta$-coordinate attains its maximum and minimum values for $\psi_p=-\pi/2$ and $\psi_p=\pi/2$, respectively. Now, The two-dimensional Cartesian coordinates for the chosen observer of the velocity four-vector $\bm{e}_0$ are obtained by applying the stereographic projection of the celestial sphere $(\psi_p,\vartheta_p)$ onto a plane. This yields the coordinates \cite{Grenzebach:2014}
\begin{eqnarray}\label{eq:Xp,Yp}
 &&   X_p = -2 \tan\left(\frac{\vartheta_p}{2}\right)\sin\psi_p,\label{eq:Xp}\\
 &&   Y_p = -2 \tan\left(\frac{\vartheta_p}{2}\right)\cos\psi_p.\label{eq:Yp}
\end{eqnarray}
The case of $\theta_o = \pi/2$ depicts an edge-on view of the shadow, while $\theta_o = 0$ represents a face-on perspective. Although the case of $\theta_o=0$ is not well-defined in the coordinate $\psi_p$ as seen in Eq. \eqref{eq:psi_p}, it is irrelevant in the context of our static black hole. In such a scenario, there would be no distinction between the edge-on and face-on views. Thus, we can proceed with the case of $\theta_o = \pi/2$ without considering this issue further. Using Eqs. \eqref{eq:Xp} and \eqref{eq:Yp}, along with the definitions in \eqref{eq:psi_p} and \eqref{eq:vartheta_p}, we can derive the expression
\begin{equation}
R_s = \sqrt{X_p^2+Y_p^2} = 2\tan\left(\frac{1}{2}\arcsin\left(\frac{\sqrt{f(r_o)\,\eta_p(r_p)}}{r_o}\right)\right).
    \label{eq:Rs}
\end{equation}
Since the black hole is static, the photon ring in the observer's sky would appear as a circle with a radius of $R_s$, as defined in the equation above.

\subsection{The energy emission rate}

At high energy levels, Hawking radiation is typically emitted within a finite cross-sectional area, denoted as $\sigma_l$. For distant observers positioned far from the black hole, this cross-section gradually approaches the shadow cast by the black hole \cite{belhaj_deflection_2020,wei_observing_2013}. It has been shown that $\sigma_l$ is directly linked to the area of the photon ring and can be mathematically represented as \cite{wei_observing_2013,decanini_fine_2011,li_shadow_2020}
\begin{equation}
\sigma_l\approx \pi R_s^2.
    \label{eq:sigmal}
\end{equation}
Accordingly, the energy emission rate of the black hole can be expressed as
\begin{equation}
\Omega\equiv\frac{\ed^2 E(\varpi)}{\ed\varpi\,\ed t} = \frac{2\pi^2\sigma_l}{e^{\varpi/T_\mathrm{H}^+}-1}\,\varpi^3,
    \label{eq:emissionrate}
\end{equation}
in which $\varpi$ is the emission frequency.

\subsection{The Deflection angle}

Gravitational lensing has emerged as a powerful tool for astrophysicists, offering a unique window into the universe. The phenomenon, caused by the bending of light around massive bodies, has facilitated numerous scientific breakthroughs since the advent of general relativity. In particular, significant progress has been made in the past two decades in developing theoretical frameworks for calculating deflection angles and predicting lensing effects on the images of astrophysical objects (see, e.g., the seminal works in Refs. \cite{virbhadra_schwarzschild_2000,virbhadra_gravitational_2002,virbhadra_time_2008,virbhadra_relativistic_2009,virbhadra_distortions_2022,virbhadra_conservation_2024,virbhadra_images_2024}). 

{Here, we present the method for calculating the weak deflection angle in non-asymptotically flat spacetimes, as outlined in Ref. \cite{Ishihara:2016}. Let us define $\Psi_\mathbb{O}$ and $\Psi_\mathbb{S}$ to denote the angles measured at the positions of the observer $\mathbb{O}$ and the source $\mathbb{S}$, respectively. These angles are calculated using the relation \cite{Ishihara:2016}.
\begin{equation}
\sin \Psi = \frac{\xi}{r}\sqrt{f(r)},
    \label{eq:Psi_0}
\end{equation}
for the spacetime \eqref{metr} when $f(r)=g(r)$. Furthermore, we define $\phi_\mathbb{O}$ and $\phi_\mathbb{S}$ to denote the longitudes of the observer and the source, respectively. Thus, the quantity $\phi_{\mathbb{OS}} = \phi_{\mathbb{O}} - \phi_\mathbb{S}$ represents the separation azimuth angle between the observer and the source. With these quantities, the light deflection angle can be calculated as 
\begin{equation}
\hat\upsilon=\phi_{\mathbb{OS}}+\left(\Psi_\mathbb{O}-\Psi_\mathbb{S}\right).
    \label{eq:hatupsilon_0}
\end{equation}
To calculate the deflection angle, it is necessary to evaluate each segment of the relation described above. In the case where the spacetime is not asymptotically flat (as with the GCJDF black hole), one must consider the finite distances between the source and the observer, since spatial infinity is not causally defined in this spacetime structure. Accordingly, let us denote the two finite distances as $r_\mathbb{O}$ and $r_\mathbb{S}$, representing the distances to the observer and the source, respectively. We assume that the light deflection occurs in the equatorial plane (planar orbits), which means $\eta = 0$ (i.e., $\Q=0$). Thus, by combining Eqs. \eqref{eq:rdot} and \eqref{eq:phidot}, we obtain
\begin{equation}
\left(\frac{\ed u}{\ed\phi}\right)^2=\mathscr{F}(u),
    \label{eq:dudphi}
\end{equation}
in which 
 \begin{equation}
 \mathscr{F}(u) = \frac{1}{\xi^2}-u^2 f(u),
     \label{eq:F(u)}
 \end{equation}
 and $u=1/r$. Accordingly, the change in the azimuth angle $\phi_{\mathbb{OS}}$ is obtained by integrating Eq. \eqref{eq:dudphi} over an appropriate range. Thus, the deflection angle in Eq. \eqref{eq:hatupsilon_0} is calculated as 
 \begin{equation}
     \hat{\upsilon} =\int_{u_\mathbb{O}}^{u_c}\frac{\ed u}{\sqrt{\mathscr{F}(u)}}+\int_{u_\mathbb{S}}^{u_c}\frac{\ed u}{\sqrt{\mathscr{F}(u)}} +\left(\Psi_\mathbb{O}-\Psi_\mathbb{S}\right),
     \label{eq:hatupsilon_1}
 \end{equation}
 with $u_c$ being the closest approach to the black hole during the lensing process.
}

In the following section, we apply the formulations introduced in this section to the GCJDF black hole to calculate the geometrical shapes of the shadow and the deflection angle. Additionally, we propose constraints on the black hole parameters based on the outcomes of the EHT.

\section{Shadow and the deflection angles of the GCJDF black hole}\label{sec:SDcalc}

In this section, we examine the shadow and deflection angle of the GCJDF black hole by employing the framework outlined in the previous section. Our objective is to analyze how various factors such as the $\mathcal{B}$-parameter influence the behavior of the shadow and deflection angle. Specifically, we apply the line element corresponding to the black hole to the derived formulas within our framework. Through this investigation, we aim to determine how the shadow behavior of the black hole is sensitive to changes in the intrinsic black hole parameters.

\subsection{The effective potential}

\begin{figure}[t]
    \centering
    \includegraphics[width=8cm]{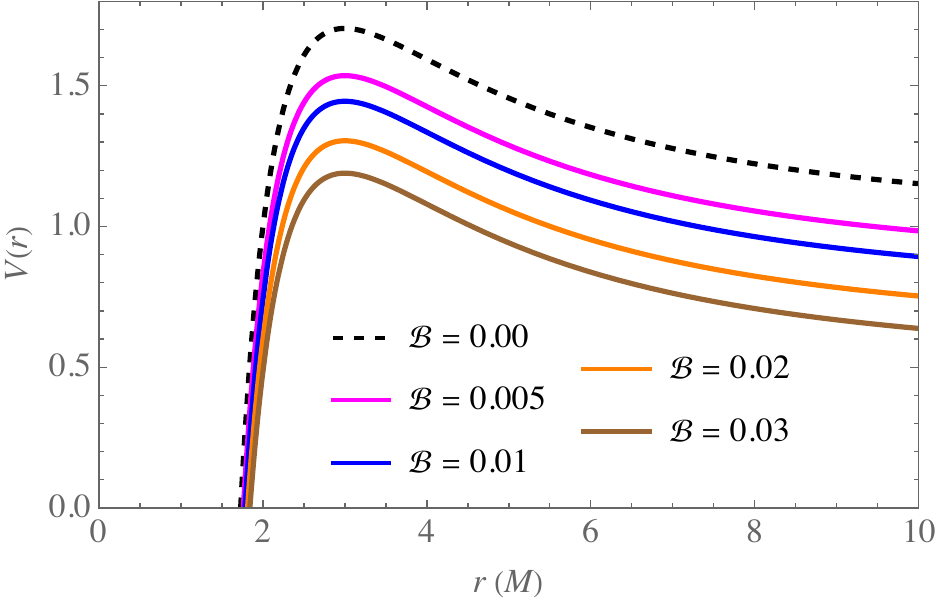}(a)\qquad
    \includegraphics[width=8cm]{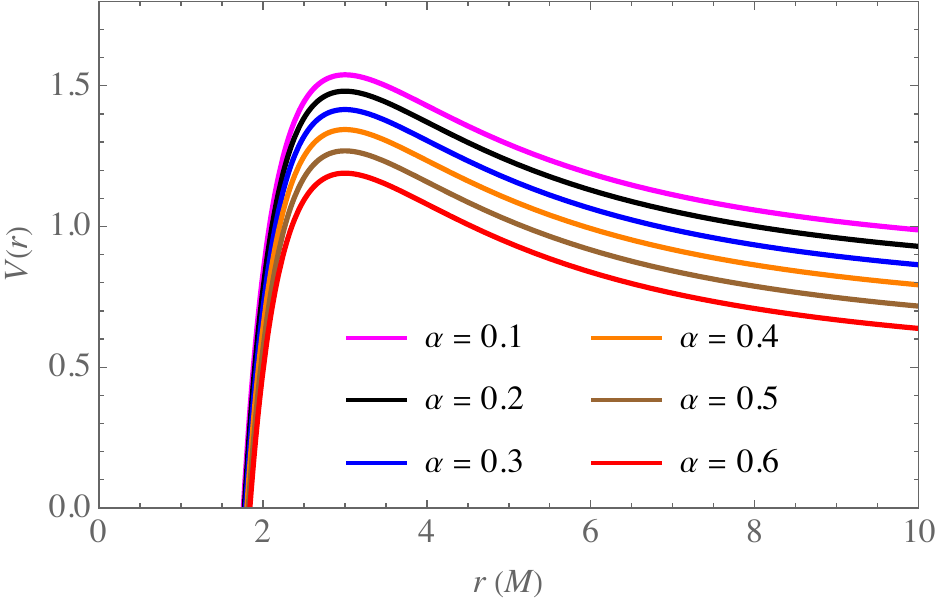}(b)
    \caption{The radial profile of the effective potential plotted for $\m=0.4$, $q=0.1$, $\rho_0=1$, $\Q=10$, $E=1$, $L=3$, and (a) five different values of the $\mathcal{B}$-parameter while $\alpha=0.6$, and (b) six different values of the $\alpha$-parameter while $\mathcal{B}=0.03$.}
    \label{fig:V_2}
\end{figure}

To obtain the exact expression for the effective potential, let us exploit Eqs. \eqref{lapf01},  \eqref{lambfunc} and  \eqref{eq:F1trans}, in Eq.  \eqref{eq:V}, which provide
\begin{multline}
V(r) = \frac{\left(L^2+\Q\right) (r-2 M)}{r^3}-E^2-\frac{1}{3} \rho_0 \left(L^2+\Q\right) \left[1-\frac{\m q^2 r^{-3 (\alpha +1)}}{\m'}\right]^{\frac{1}{\alpha +1}} \\
\times \left[\frac{\m' \mathcal{B} \left(q^2+r^{3 \alpha +3}\right) \left(\m q^2-\m' r^{3 \alpha +3}\right)}{\m q^2 r^{3 \alpha +3}+\m' r^{6 \alpha +6}}\right]^{\frac{1}{\alpha +1}} 
{}_2F_1\left(1,-\frac{1}{\alpha +1};\frac{\alpha }{\alpha +1};\frac{1}{\m-\frac{\m' r^{3 \alpha +3}}{q^2}}\right).
    \label{eq:V_2}
\end{multline}
In Fig. \ref{fig:V_2}, the radial profile of the effective potential has been plotted for different values of the black hole parameters. 
Observing the diagrams, it is evident that variations in the $\mathcal{B}$-parameter yield more noticeable effects compared to alterations in the $\alpha$-parameter, as indicated by the greater separation between the peaks of the effective potential.
From the right panel of the figure, we observe that an increase in the $\alpha$-parameter lowers the peak of the effective potential. This indicates that as the dark fluid becomes {less} anisotropic, the black hole's tendency to bend light decreases, resulting in fewer deflecting trajectories. This phenomenon will be further emphasized in Subsect. \ref{subsec:delfangle}, where the deflection angles are analyzed.
Additionally, it is noteworthy that the Schwarzschild black hole dominates the highest photon energies, given its larger effective potential's peak. Notably, alterations in other spacetime parameters do not significantly alter the overall behavior of the effective potential. Particularly, the parameter $\rho_0$ serves as a scaling factor; lower values of $\rho_0$ correspond to higher peaks in the effective potential. Nonetheless, the Schwarzschild black hole consistently serves as an upper limit across all cases of the GCJDF black hole.

\subsection{Black hole shadow}

\allowdisplaybreaks

\begin{figure}[t]
    \centering
    \includegraphics[width=5.3cm]{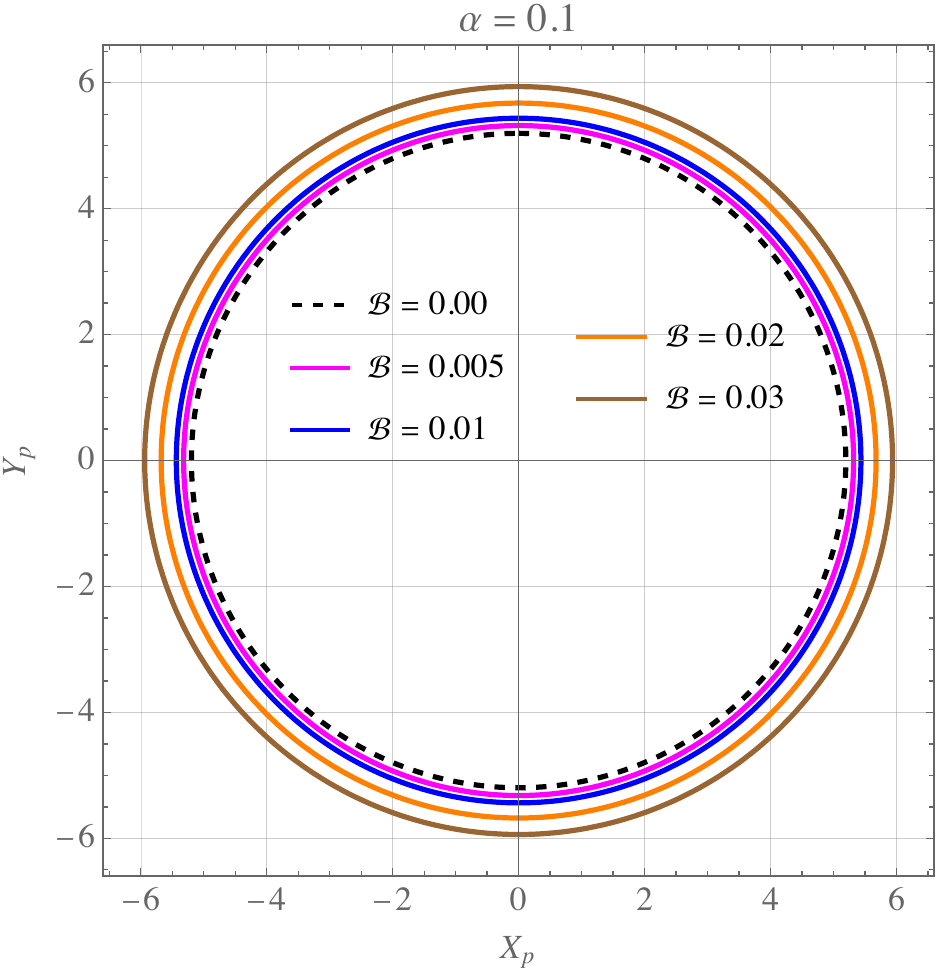}(a)\quad
    \includegraphics[width=5.3cm]{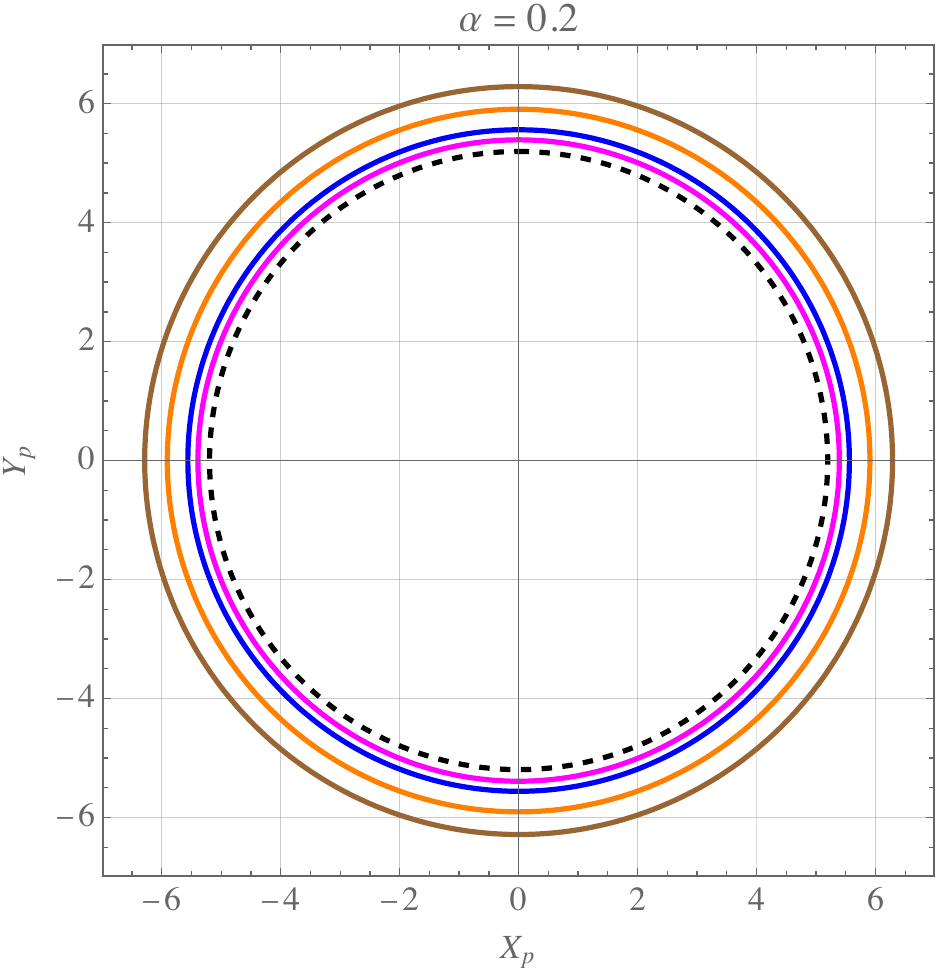}(b)\quad
    \includegraphics[width=5.3cm]{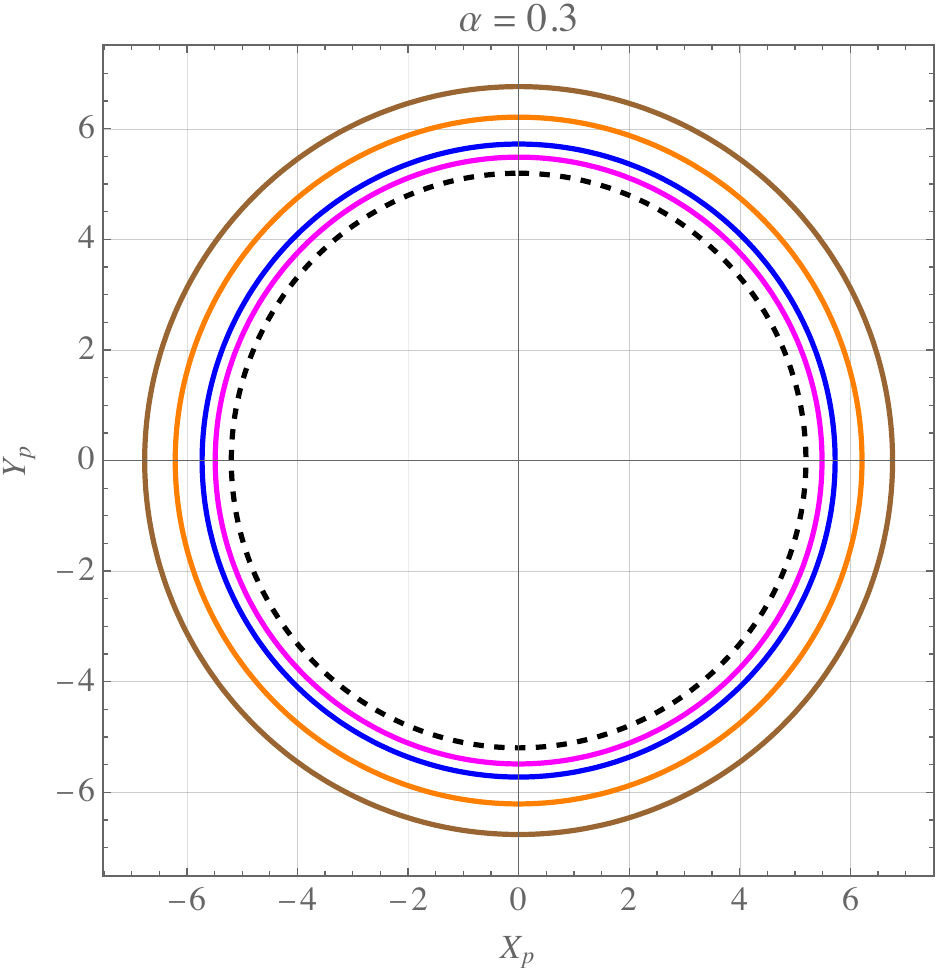}(c)
    \includegraphics[width=5.3cm]{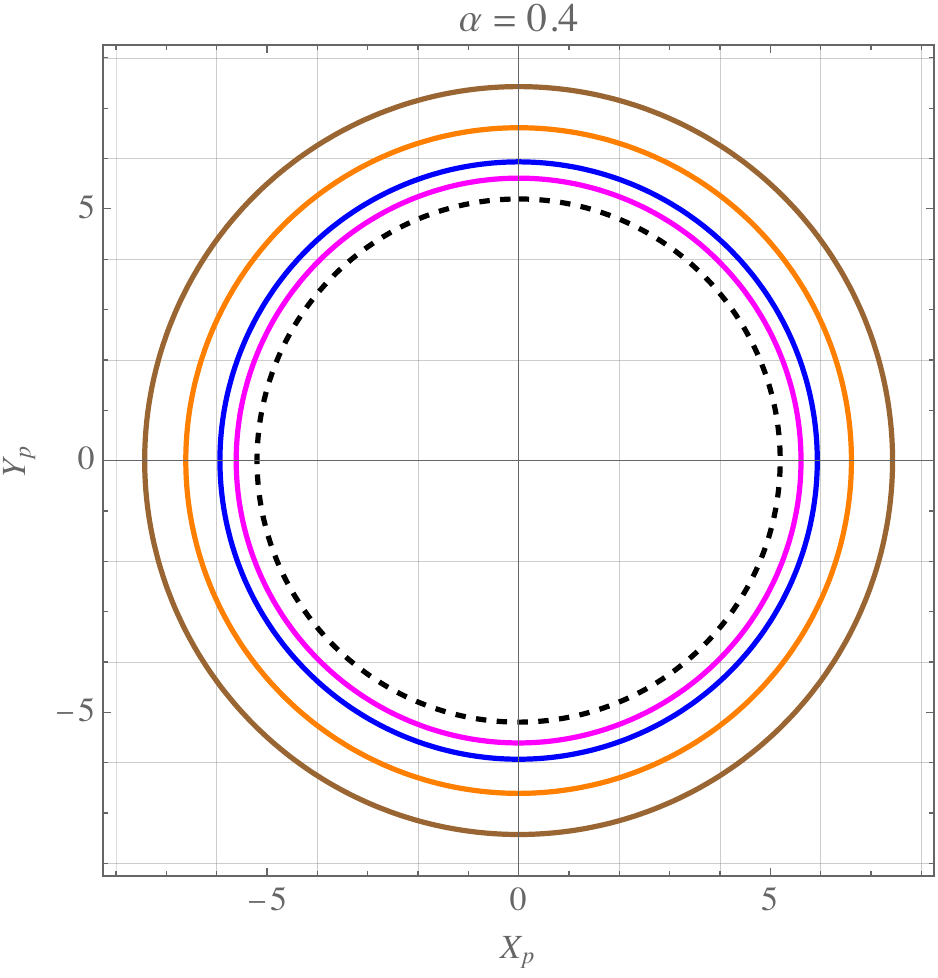}(d)\quad
    \includegraphics[width=5.3cm]{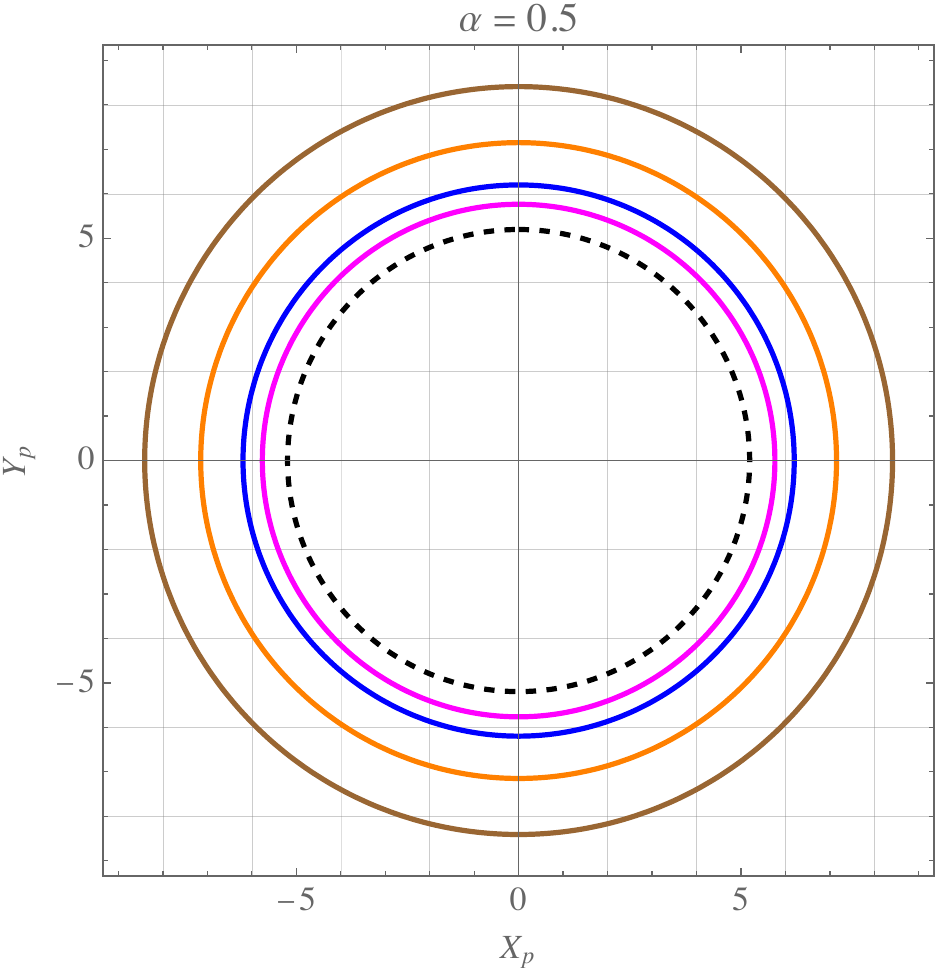}(e)\quad
    \includegraphics[width=5.3cm]{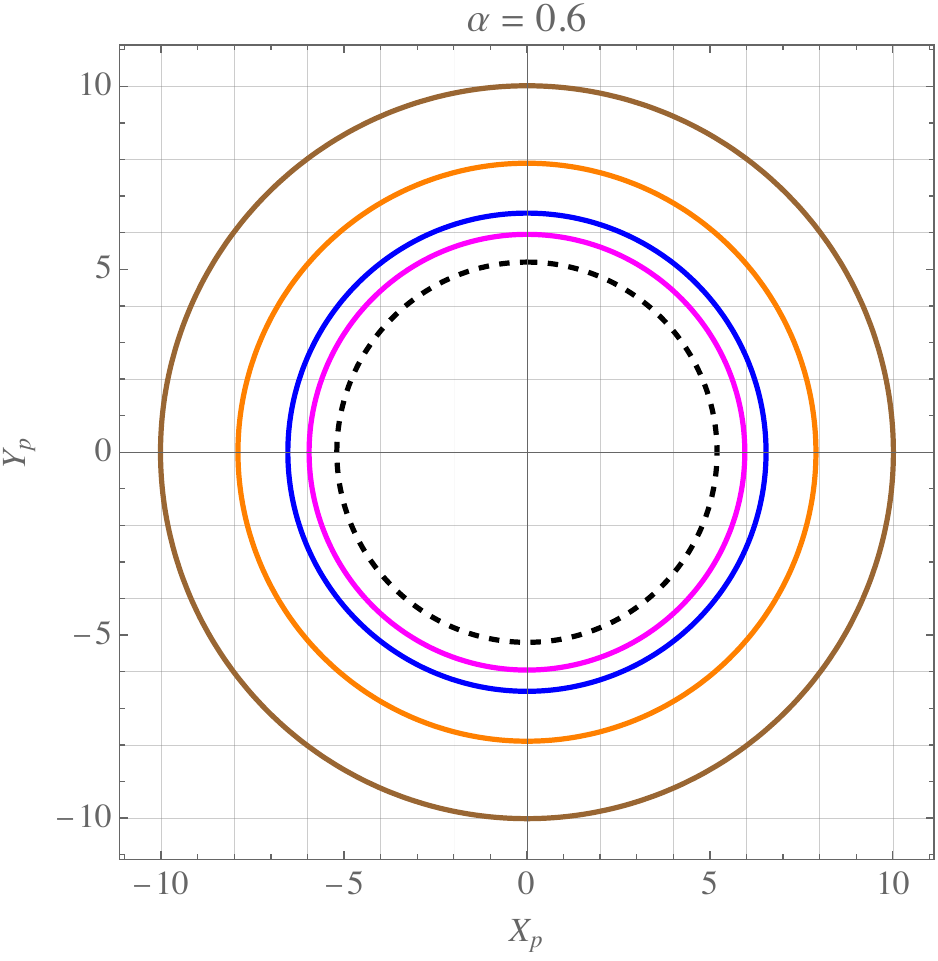}(f)
    \caption{The circles corresponding to the critical curves for the GCJDF black hole with the intrinsic parameters $\m=0.4$, $q=0.1$ and $\rho_0=1$, plotted by considering an observer located at $(100,\pi/2)$. Each panel corresponds to a different value of the $\alpha$-parameter and the curves indicate the photon ring for a specific value for the $\mathcal{B}$-parameter.}
    \label{fig:shadow}
\end{figure}

Applying the formulation provided in Subsect. \ref{subsec:shadowparam}, we can characterize the outer boundary of the black hole shadow, referred to as the \textit{critical curve}. To do this, we utilize the impact parameter from Eq. \eqref{eq:etap} for specific values of the black hole parameters, enabling the calculation of the shadow radius $R_s$ in Eq. \eqref{eq:Rs}. In Table \ref{tab:2}, some exemplary values of $r_+$, $r_p$ and $R_s$ are provided for various cases of intrinsic black hole parameters.
\begin{table}[htp]
        \centering
        \begin{tabular}{|c||c|c|c||c|c|c||c|c|c||c|c|c||c|c|c|}
        \hline
             & \multicolumn{3}{|c||}{$\mathcal{B}=0.00$} & \multicolumn{3}{|c||}{$\mathcal{B}=0.005$} & \multicolumn{3}{|c||}{$\mathcal{B}=0.01$} & \multicolumn{3}{|c||}{$\mathcal{B}=0.02$} & \multicolumn{3}{|c|}{$\mathcal{B}=0.03$}\\
             \cline{2-16}
          $\alpha$   & $r_{+}$ & $r_{p}$ & $R_{s}$ &  $r_{+}$ & $r_{p}$ & $R_{s}$ & $r_{+}$ & $r_{p}$ & $R_{s}$ & $r_{+}$ & $r_{p}$ & $R_{s}$ & $r_{+}$ & $r_{p}$ & $R_{s}$ \\
            \hline\hline
            \multicolumn{1}{|c||}{$0.1$} & 2.00 & 3.00 & 5.196 &  2.014 & 3.00032 & 5.320 &  2.023 & 3.0006 & 5.440 & 2.052 & 3.00112 & 5.680 &  2.079 & 3.00162 & 5.940\\
            \hline
            \multicolumn{1}{|c||}{$0.2$} &  2.00 & 3.00 & 5.196 &  2.022 & 3.00018 & 5.392 &  2.040 & 3.00033 & 5.561 &  2.075 & 3.00058 & 
            5.910 &  2.111 & 3.00082 & 6.290\\
            \hline
            \multicolumn{1}{|c||}{$0.3$} &  2.00 & 3.00 & 5.196 &  2.032 & 3.00013 & 5.487 &  2.057 & 3.00023 & 5.724 &  2.104 & 3.00039 & 6.210 &  2.151 & 3.00053 & 6.763\\
            \hline
            \multicolumn{1}{|c||}{$0.4$} &  2.00 & 3.00 & 5.196 &  2.045 & 3.0001 & 5.610 &  2.077 & 3.00017 & 5.933 &  2.139 & 3.00027 & 6.611 &  2.202 & 3.00037 & 7.430\\
            \hline
            \multicolumn{1}{|c||}{$0.5$} &  2.00 & 3.00 & 5.196 &  2.061 & 3.00008 & 5.764 &  2.102 & 3.00013 & 6.198 &  2.182 & 3.0002 & 7.150 &  2.267 & 3.00026 & 8.410\\
            \hline
            \multicolumn{1}{|c||}{$0.6$} & 2.00 & 3.00 & 5.196 &  2.079 & 3.00006 & 5.955 & 2.132 & 3.0001 & 6.540 & 2.234 & 3.00015 & 7.900
            & 2.353 & 3.00019 & 10.016\\
            \hline
            \hline
        \end{tabular}
        \caption{Values of $r_{+}$, $r_{p}$, and $R_s$ for the GCJDF black hole at different values of $\mathcal{B}$ and $\alpha$, with $\m=0.4$, $q=0.1$, and $\rho_0=1$.}
        \label{tab:2}
\end{table}
These data are subsequently utilized in Fig. \ref{fig:shadow} to visually represent the geometric shapes of the GCJDF black hole shadow.
As we can observe from the diagrams, for fixed values of $\alpha$, a smaller $\mathcal{B}$-parameter results in a smaller shadow diameter. Accordingly, the Schwarzschild black hole has the smallest shadow, constituting the lower limit regarding the shadow size. Furthermore, for fixed $\mathcal{B}$, a larger $\alpha$-parameter results in a larger shadow size. This means that {less}-anisotropic dark fluids result in bigger black hole shadows in the observer's sky. Moreover, for larger $\alpha$, the difference in the shadow size for different values of the $\mathcal{B}$-parameter is increased; the corresponding critical curves are more separated.

\raggedbottom

\subsection{Energy emission rate}

\begin{figure}[h!]
    \centering
    \includegraphics[width=5.3cm]{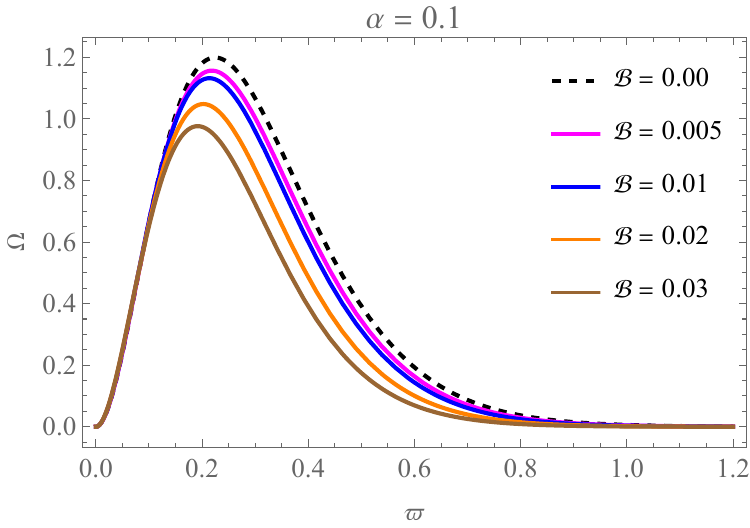}(a)\quad
    \includegraphics[width=5.3cm]{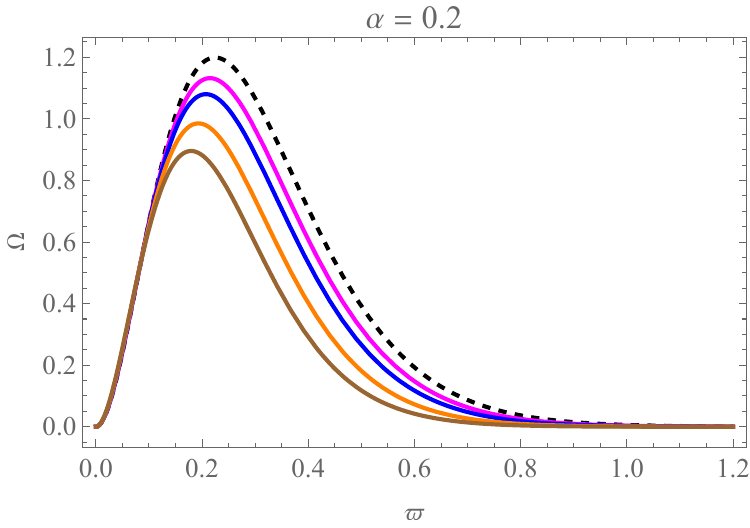}(b)\quad
    \includegraphics[width=5.3cm]{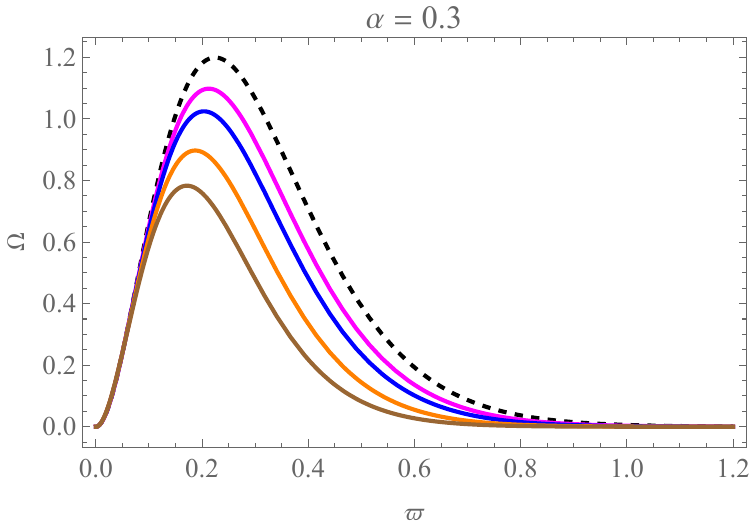}(c)
    \includegraphics[width=5.3cm]{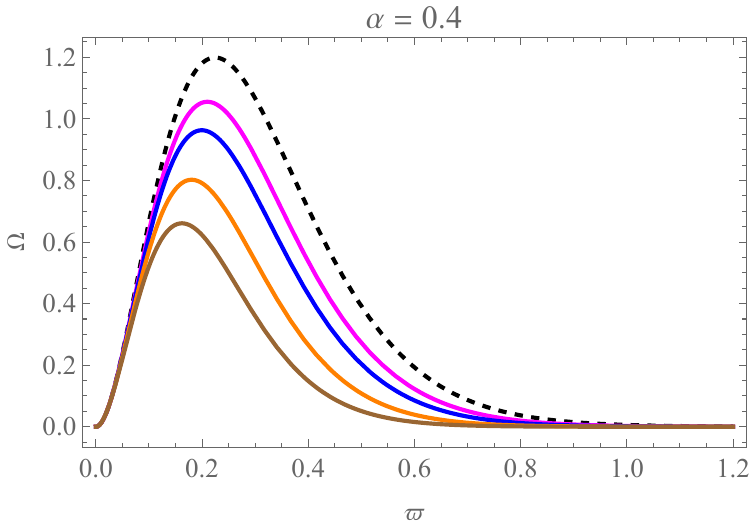}(d)\quad
    \includegraphics[width=5.3cm]{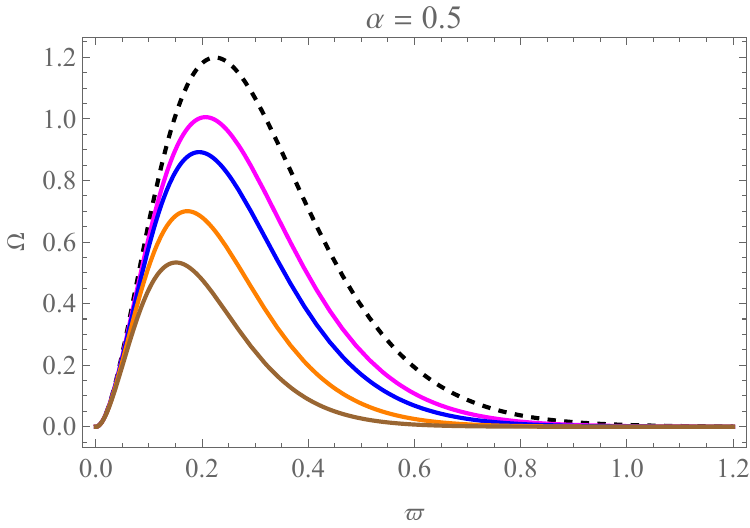}(e)\quad
    \includegraphics[width=5.3cm]{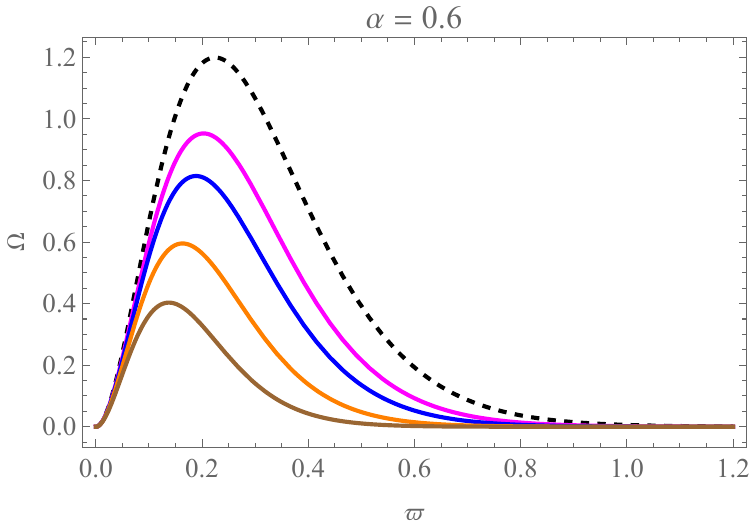}(f)
    \caption{The $\varpi$-profile of the emission frequency plotted for the data given in Table \ref{tab:2}.}
    \label{fig:Omega}
\end{figure}

Similar to what was done in Sect. \ref{sec:AdSBH_sol}, to find the numerical values of the Hawking temperature of the black hole, we can now use the $r_{+}$ values from Table \ref{tab:2}. Next, the numerical values of $\sigma_{l}$ in Eq. \eqref{eq:sigmal} for different $\mathcal{B}$ and $\alpha$ should be determined from the shadow radius values, $R_s$, of the black hole. Consequently, by inserting the Hawking temperature and $\sigma_{l}$ values into Eq. \eqref{eq:emissionrate}, one can find the expressions for the energy emission rate in terms of different $\mathcal{B}$ and $\alpha$ associated with the GCJDF black hole. In Fig. \ref{fig:Omega}, the profile of the energy emission rate has been plotted as a function of the emission frequency. 
As observed from the diagrams, for all fixed values of the $\alpha$-parameter, the energy emission rate for the $\mathcal{B}=0$ case constitutes the upper limit; the Schwarzschild black hole evaporates faster, and an increase in $\mathcal{B}$ decelerates the evaporation process. Conversely, an increase in the $\alpha$-parameter, when $\mathcal{B}$ is fixed, results in lower rates of energy emission. Thus, the {less} anisotropic the dark fluid, the less inclined the black hole is to evaporation. Furthermore, for larger $\alpha$, the differences in the energy emission rate for different values of the $\mathcal{B}$-parameter become more noticeable.

\subsection{Deflection angle}\label{subsec:delfangle}

\begin{figure}[t]
    \centering
    \includegraphics[width=5.3cm]{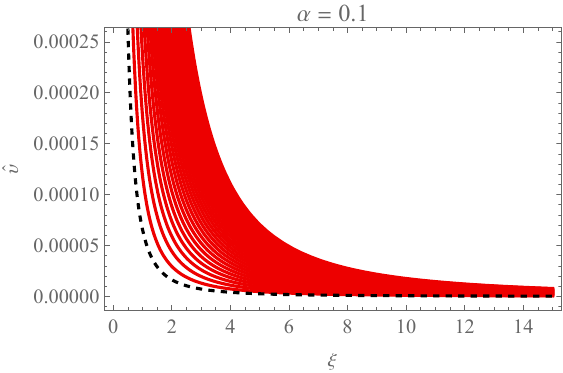}(a)\quad
    \includegraphics[width=5.3cm]{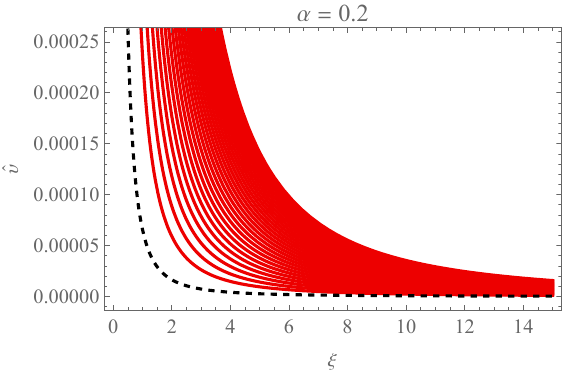}(b)\quad
    \includegraphics[width=5.3cm]{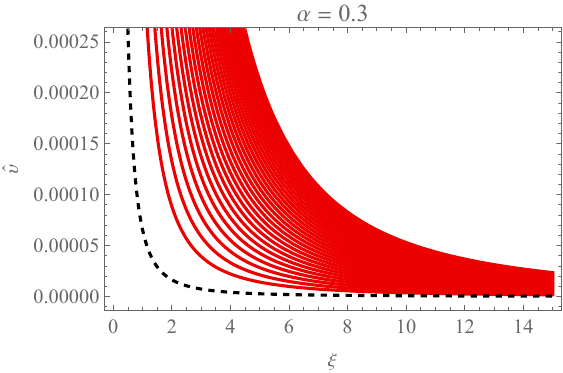}(c)
    \includegraphics[width=5.3cm]{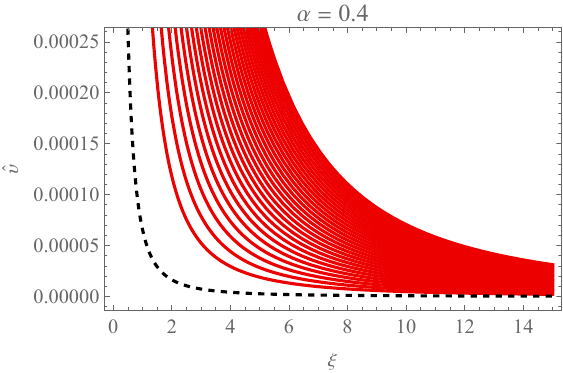}(d)\quad
    \includegraphics[width=5.3cm]{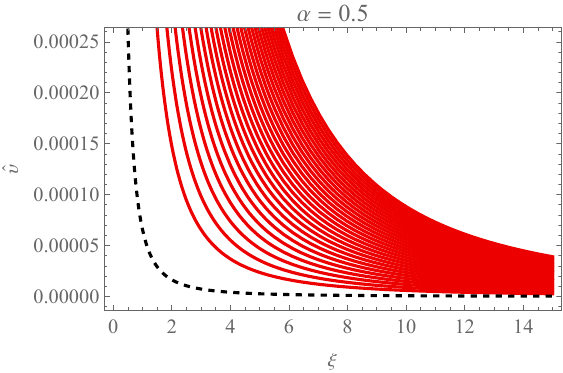}(e)\quad
    \includegraphics[width=5.3cm]{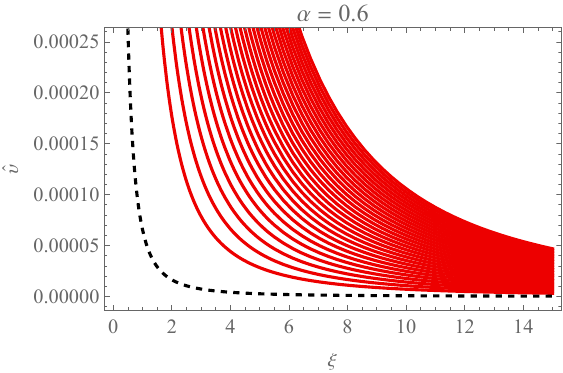}(f)
    \caption{{The behavior of the deflection angle $\hat\upsilon$ in arcseconds as a function of the impact parameter $\xi$, plotted for the intrinsic parameters $\m=0.4$, $q=0.1$, and $\rho_0=1$, with the observer and the source located at $r_\mathbb{O}=100$ and $r_\mathbb{S}=100$. Each panel corresponds to a specific value of the $\alpha$-parameter, and the curves indicate the behavior of the deflection angle for a range of the $\mathcal{B}$-parameter within the domain $\mathcal{B}\in(0,0.03)$, with increments of $0.001$. The dashed curve represents the case of $\mathcal{B}=0$.}}
    \label{fig:deflection}
\end{figure}

{Applying the geometry of the GCJDF black hole, the characteristic polynomial in Eq. \eqref{eq:F(u)} can be approximated as
\begin{equation}
\mathscr{F}(u)=a u^3-u^2+b u +c,
    \label{eq:F(u)_1}
\end{equation}
up to the first order in $\alpha$ and $\mB$, where 
\begin{subequations}
    \begin{align}
        & a =2M+\frac{1}{6} (3 \alpha +2) q^2 \rho_0 \mathcal{B} \ln \m'-\frac{(3 \alpha +2) q^2 \rho_0 \mathcal{B}}{6 \alpha },\\
        & b =-\frac{1}{2} q^2 \rho_0 \mathcal{B} \bigl(\alpha  \ln \m'-1\bigr),\\
        & c = \frac{1}{\xi^2}.
    \end{align}
    \label{eq:a,b,c}
\end{subequations}
For $\mB=0$, this polynomial correctly reduces to $\mathscr{F}(u)=1/\xi^2-u^2+2M u^3$, corresponding to the Schwarzschild black hole. By applying the additional change of variable $u=4/a\left(U+1/12\right)$, the two included integrals in Eq. \eqref{eq:hatupsilon_1} are calculated analytically as
\begin{eqnarray}
    && \phi_\mathbb{O} = \int_{U_\mathbb{O}}^{U_c}\frac{\ed U}{\sqrt{\mathscr{F}_3(U)}} = \wp^{-1}(U_\mathbb{O})-\wp^{-1}(U_c),\\
    && \phi_\mathbb{S} = \int_{U_c}^{U_\mathbb{S}}\frac{\ed U}{\sqrt{\mathscr{F}_3(U)}} = \wp^{-1}(U_c)-\wp^{-1}(U_\mathbb{S}),
\end{eqnarray}
in which $\mathscr{F}_3(U)=4U^3-g_2 U-g_3$, where 
\begin{subequations}
    \begin{align}
        & g_2 = \frac{a^2}{16}  \left(\frac{4}{3 a^2}-\frac{4 b}{a}\right),\\
        & g_3 =-\frac{a^2}{16} \left(-\frac{2}{27 a^2}+\frac{b}{3 a}+c\right),
    \end{align}
    \label{eq:g23}
\end{subequations}
and $\wp^{-1}(x)\equiv \wp^{-1}(x;g_2,g_3)$ is the inverse Weierstrassian elliptic $\wp$ function with the invariants $g_2$ and $g_3$ \cite{handbookElliptic}. Furthermore, since the lensing process involves deflecting trajectories in the equatorial plane, the quantity $U_c$ is thus the smallest root of the equation $\mathscr{F}_3(U)=0$, which is obtained as 
\begin{equation}
U_c = \sqrt{\frac{g_2}{3}} \cos \left(\frac{1}{3} \arccos \left(\frac{3g_3}{g_2}\sqrt{\frac{3}{g_2}}\,\right)-\frac{2 \pi }{3}\right).
    \label{eq:uc}
\end{equation}
The remaining components of the lens equation \eqref{eq:hatupsilon_1}, can be calculated by means of Eq. \eqref{eq:Psi_0}. Accordingly, after algebraic manipulation, the weak deflection angle up to the first order in $\alpha$ and $\mB$ is obtained as 
\begin{multline}
\hat\upsilon = \wp^{-1}(U_\mathbb{O})+\wp^{-1}(U_\mathbb{S})-2\wp^{-1}(U_c)
+\arcsin\left( \xi  U_\mathbb{O} \sqrt{2 M U_\mathbb{O}}\right)-\arcsin\left( \xi  U_\mathbb{S} \sqrt{ 2 M U_\mathbb{S}}\right)
\\
+\frac{1}{12 \alpha  M U_\mathbb{O} U_\mathbb{S} \sqrt{4 M \xi ^2 U_\mathbb{O}^3+2} \sqrt{2 M \xi ^2 U_\mathbb{S}^3+1}}\Biggl\{
(\alpha +1) \xi  q^2 \mathcal{B}\rho_0\bigl(1-\alpha\ln\m'\bigr)\Biggl[
U_\mathbb{O} \sqrt{M U_\mathbb{S}} \bigl[(3 \alpha +2) U_\mathbb{S}^2-3 \alpha \bigr] \sqrt{2 M \xi ^2 U_\mathbb{O}^3+1}
\\
+3 \alpha  U_\mathbb{S} \sqrt{M U_\mathbb{O}} \sqrt{2 M \xi ^2 U_\mathbb{S}^3+1}-(3 \alpha +2) U_\mathbb{O}^2 U_\mathbb{S} \sqrt{M U_\mathbb{O}} \sqrt{2 M \xi ^2 U_\mathbb{S}^3+1}
\Biggr]
\Biggr\}
+\mathcal{O}(\alpha^2,\mB^2).
    \label{eq:hatupsilon_2}
\end{multline}
Taking into account the performed changes of variables, Fig. \ref{fig:deflection} shows the $\xi$-profile of the deflection angle presented in Eq. \eqref{eq:hatupsilon_2}. The profiles have been plotted for different values of the $\alpha$-parameter and within a certain domain for the $\mB$-parameter.} 
As inferred from the diagrams, an increase in the $\mathcal{B}$-parameter, while keeping $\alpha$ fixed, results in the same values of the deflection angle being obtained for larger impact parameters. A similar behavior is observed for the $\mathcal{B}$-constant curves when the $\alpha$-parameter is increased; the same values of the deflection angle are achieved for larger impact parameters. Consequently, the curves fall less steeply for higher values of $\alpha$. Thus, when the fluid is {less} anisotropic, the black hole causes less gravitational lensing and bends light to a lesser extent. This aligns with our expectations regarding the sensitivity of the effective potential to variations in the $\alpha$-parameter, as shown in Fig. \ref{fig:V_2}(b).

\subsection{Observational constraints from the EHT}\label{subsec:constraints}

\begin{figure}[t]
    \centering
    \includegraphics[width=5.3cm]{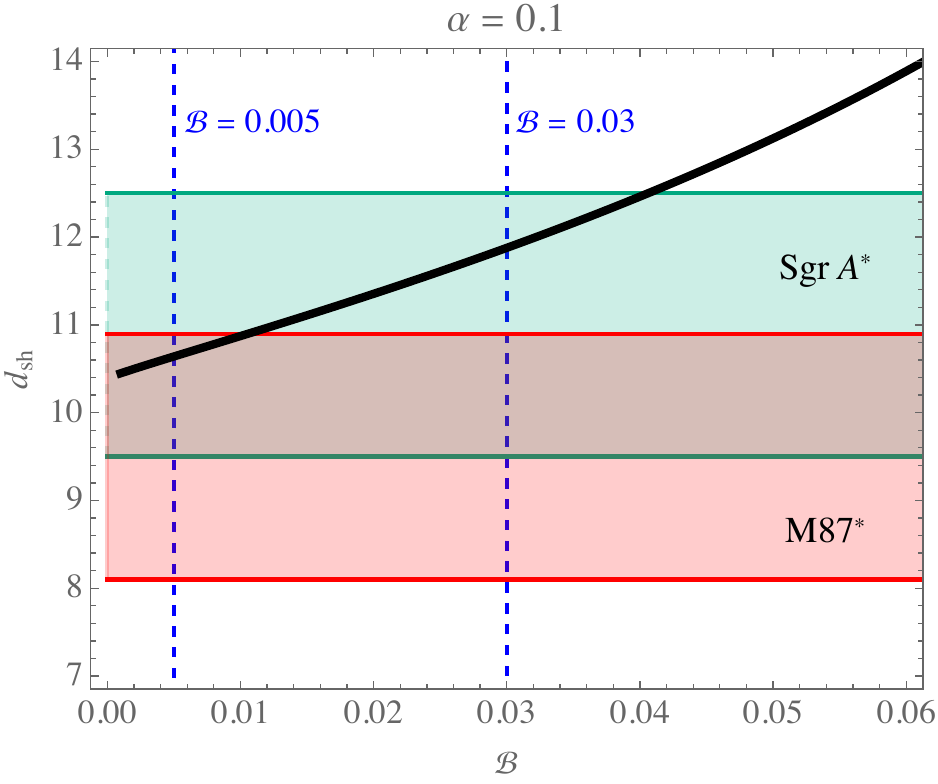}(a)\quad
    \includegraphics[width=5.3cm]{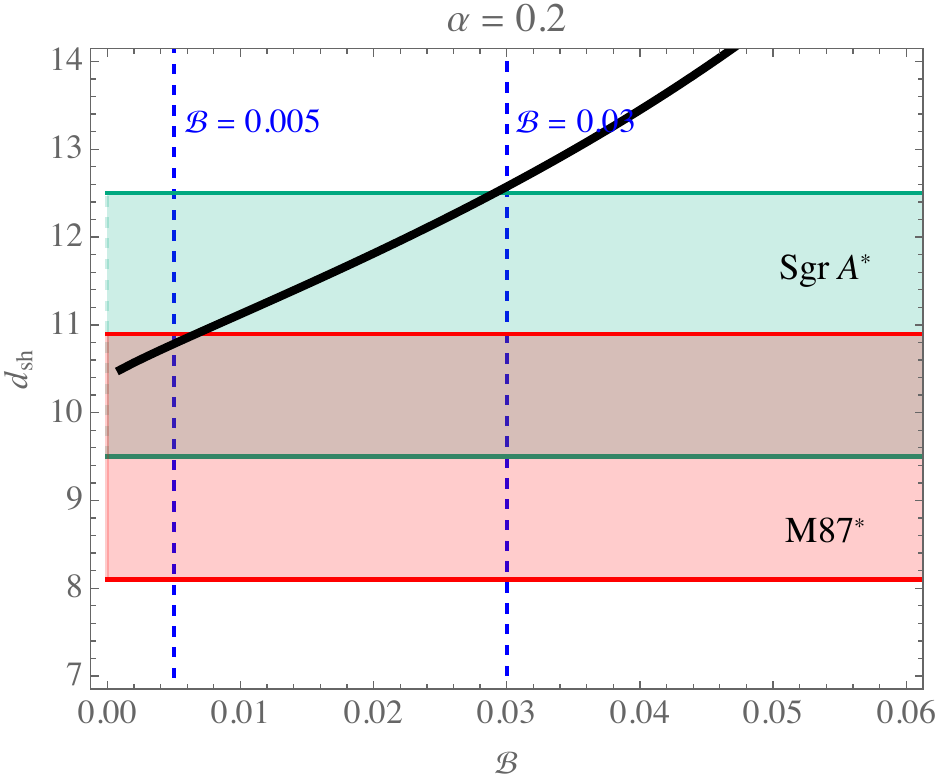}(b)\quad
    \includegraphics[width=5.3cm]{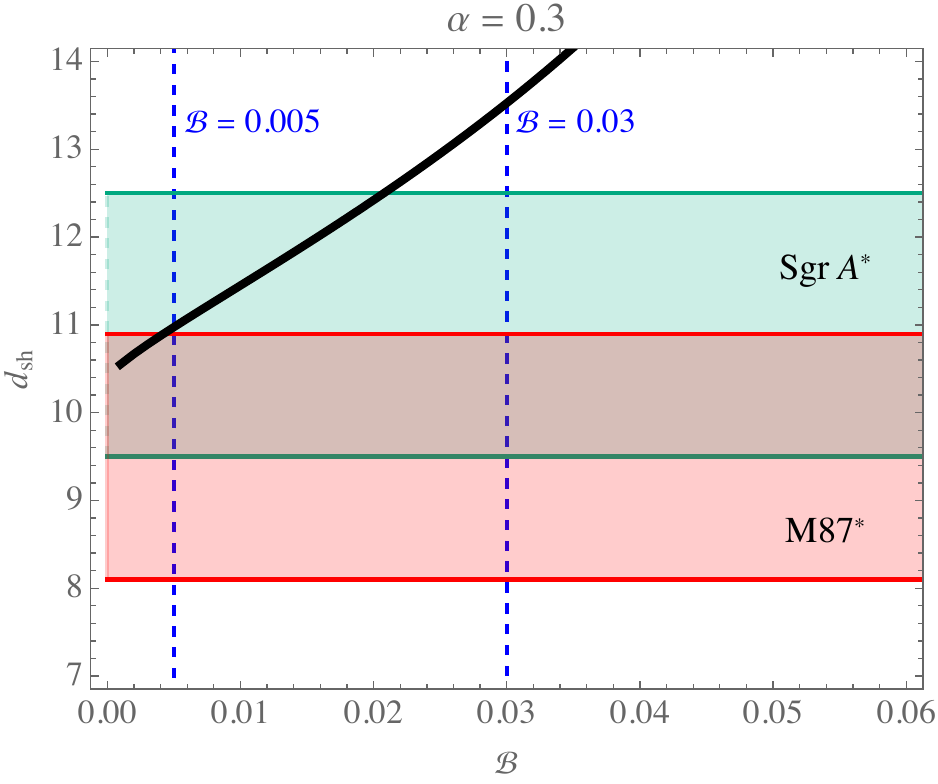}(c)
    \includegraphics[width=5.3cm]{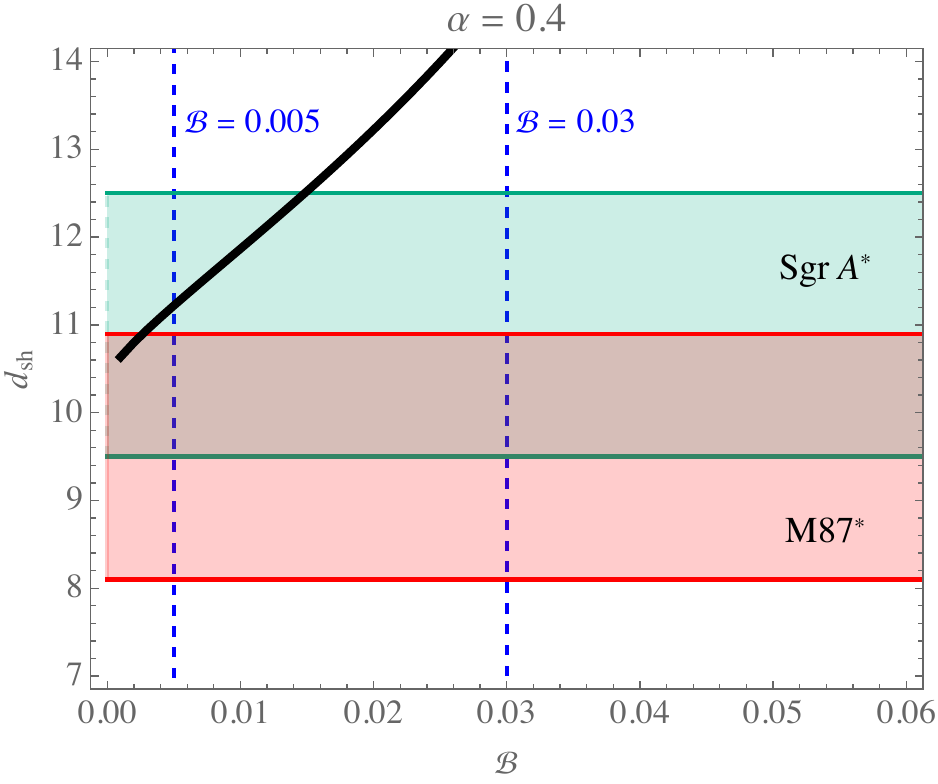}(d)\quad
    \includegraphics[width=5.3cm]{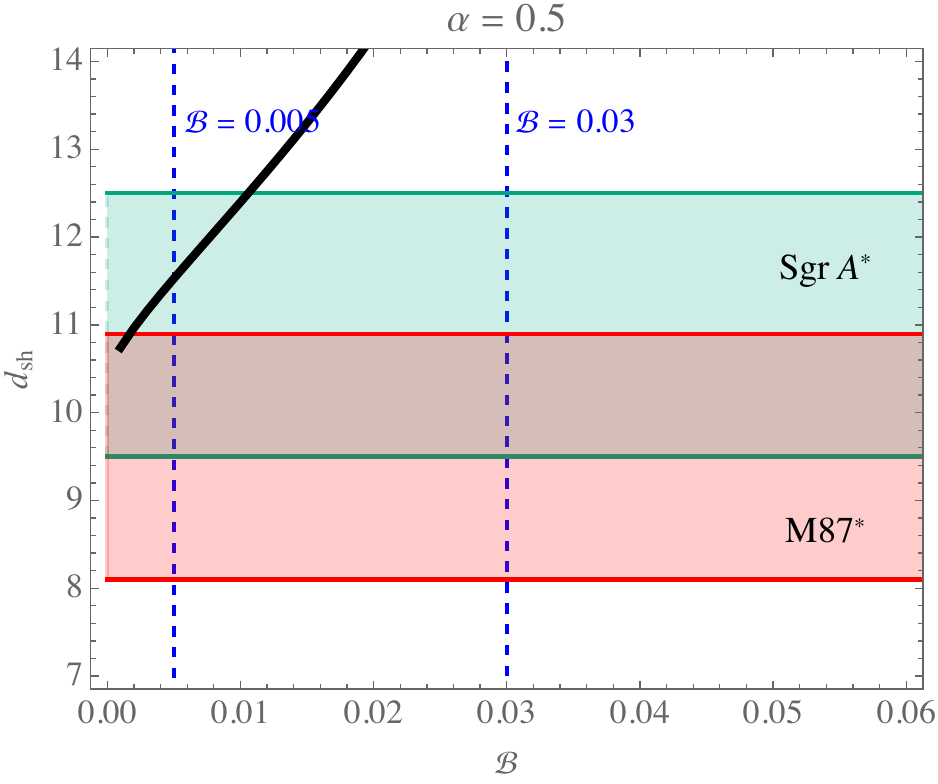}(e)\quad
    \includegraphics[width=5.3cm]{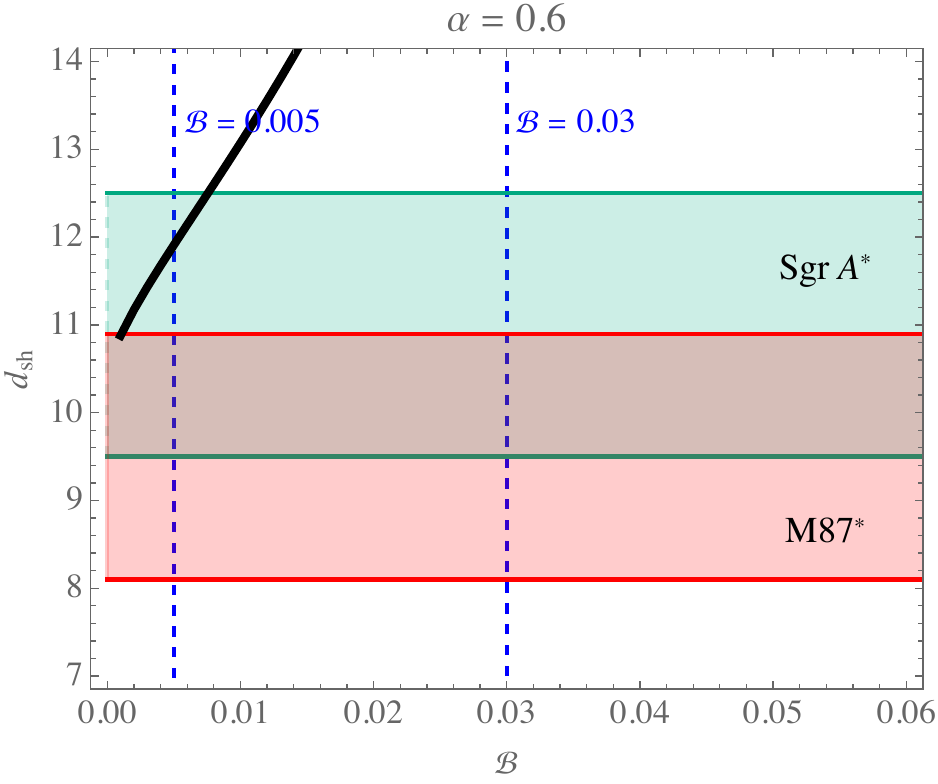}(f)
    \caption{The $\mB$-profiles of the theoretical shadow diameter $d_{\mathrm{sh}}^{\mathrm{theo}}$ (black curves) compared with the observed shadow diameters of M87* and Sgr A*, for various values of the $\alpha$-parameter. The plots are generated for $\m=0.4$, $q=0.1$, and $\rho_0=1$.
    }
    \label{fig:constraints}
\end{figure}

To enable comparisons with the EHT observations, we recall that the theoretical shadow diameter for the black hole is given by $d_{\mathrm{sh}}^{\mathrm{theo}}=2R_{s}$, where $R_s$ is defined in Eq. \eqref{eq:Rs}. To calculate the diameter of the observed shadows in the recent EHT images of M87* and Sgr A*, we use the relation \cite{Bambi:2019tjh}
\begin{equation}
    d_\text{sh} = \frac{D \theta_*}{\gamma M_\odot},    \label{eq:dsh}
\end{equation}
which calculates the shadow diameter as observed by an observer positioned at a distance $D$ (in parsecs) from the black hole. Here, $\gamma$ is the mass ratio of the black hole to the Sun. For M87*, $\gamma = (6.5 \pm 0.90) \times 10^9$ at a distance $D = 16.8\,\mathrm{Mpc}$ \cite{Akiyama:2019}, and for Sgr A*, $\gamma = (4.3 \pm 0.013) \times 10^6$ at $D = 8.127\,\mathrm{kpc}$ \cite{Akiyama:2022}. In Eq.~\eqref{eq:dsh}, $\theta_*$ is the angular diameter of the shadow, measured as $\theta_* = 42 \pm 3 \,\mathrm{\mu as}$ for M87* and $\theta_* = 48.7 \pm 7 \,\mathrm{\mu as}$ for Sgr A*. Using these values, the shadow diameters can be calculated as $d_{\mathrm{sh}}^{\mathrm{M87*}} = 11 \pm 1.5$ and $d_{\mathrm{sh}}^{\mathrm{SgrA*}} = 9.5 \pm 1.4$. These values are displayed with $1\sigma$ uncertainties for both black holes in Fig. \ref{fig:constraints}, together with the $\mB$-profile of the theoretical shadow diameter $d_{\mathrm{sh}}^{\mathrm{theo}}$. To obtain this profile, we have used Eqs. \eqref{eq:rpEq} and \eqref{eq:Rs} to produce a set of $(\mB,2R_s)$ points. 
According to the diagrams, as the $\alpha$-parameter increases, the profile tends to exit the observationally respected region. Interestingly, the initial assumed domain for the $\mathcal{B}$-parameter, which was used to generate the figures in this study, remains valid within the observational region. Based on the diagrams, it is inferred that the profile in Fig. \ref{fig:constraints}(a) exhibits the most conformity with the observations. Thus, in the context of the chosen values for the other black hole parameters, the reliable range for the parameters $\alpha$ and $\mathcal{B}$ is $0<\alpha\lesssim 0.1$ and $0\leq\mathcal{B}\lesssim 0.03$, in accordance with the EHT observations.

{Recently, the GCJDF has been compared with observational data from SN Ia, Cosmic Chronometers, and Fast Radio Bursts, leading to interesting conclusions \cite{Fortunato:2024uut}. Notably, the results for $B < 0$ align well for $\alpha > 0$, albeit with significant uncertainty. Future data will help clarify the compatibility of these findings.}

\subsection{Einstein rings}

\begin{figure}[t]
    \centering
    \includegraphics[width=9cm]{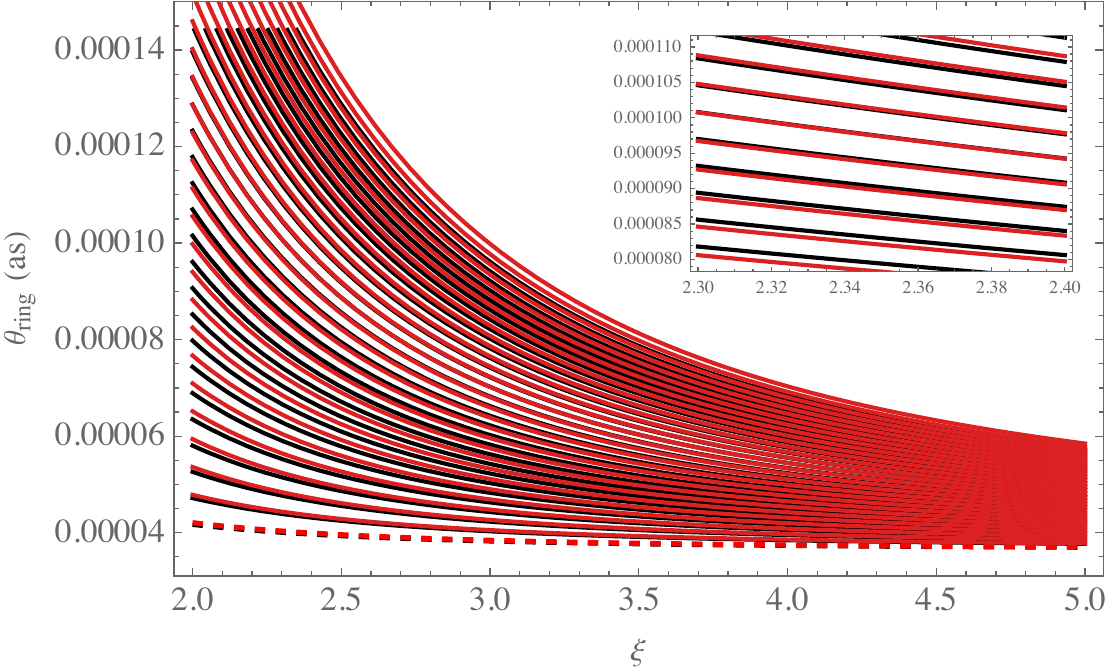}
    \caption{The profile of the Einstein ring's radius in arcseconds, plotted for M87* (black curves) and Sgr A* (red curves). The plots are generated for $\alpha=0.1$, $\m=0.4$, $q=0.1$, and $\rho_0=1$. The observer's location is considered to be $r_\mathbb{O}=D_\mathrm{R}$, and for simplicity, it is assumed that $D_\mathrm{R}=D_\mathrm{S}$ for both cases. The plots cover a range of the $\mathcal{B}$-parameter within the domain $\mathcal{B}\in(0,0.03)$, with increments of $0.001$. The dashed curves correspond to the Schwarzschild black hole for both cases.}
    \label{fig:EinRing}
\end{figure}

One of the interesting features of gravitational lensing due to light deflection is the formation of Einstein rings. In this subsection, we calculate the angular size of the Einstein rings formed by the black hole and examine the impact of the GCJDF and its anisotropies. Let us define the quantities $D_\mathrm{S}$ and $D_\mathrm{R}$ as the distances from the black hole (i.e., the lens) to the source and the receiver, respectively. Assuming a thin lens, the distance between the receiver and the source can be expressed as $D_{\mathrm{RS}}=D_{\mathrm{R}}+D_\mathrm{S}$. We then employ the lens equation \cite{Bozza:2008}
\begin{equation}
D_{\mathrm{RS}}\tan\beta = \frac{D_\mathrm{R}\sin\theta-D_\mathrm{S}\sin\left(\hat\upsilon-\theta\right)}{\sin\left(\hat\upsilon-\theta\right)},
    \label{eq:lensEq}
\end{equation}
to obtain the position of the images, in which $\hat\upsilon$ is the deflection angle calculated in Subsect. \ref{subsec:delfangle}. Einstein rings are formed when $\beta = 0$. Hence, the angular radius of the ring can be calculated from Eq. \eqref{eq:lensEq} as
\begin{equation}
\theta_{\mathrm{ring}} \approx \frac{D_\mathrm{S}}{D_\mathrm{RS}}\hat\upsilon.
    \label{eq:EinRing_eq}
\end{equation}
When the ring is small, we can further approximate by letting $\xi = D_\mathrm{R}\sin\theta \approx D_\mathrm{R}\theta$. Accordingly, the angular radius of the Einstein ring formed by the GCJDF black hole can be calculated using Eq. \eqref{eq:hatupsilon_2}, which yields
\begin{equation}
\theta_{\mathrm{ring}}^{\mathrm{GCJDF}} = \left[\frac{\mathfrak{A} D_\mathrm{S}+\sqrt{\frac{D_\mathrm{S}}{D_\mathrm{R}} \bigl(\mathfrak{A}^2 D_\mathrm{S}+{12 \mathfrak{B} D_\mathrm{RS}}\bigr)}}{6D_\mathrm{R} D_\mathrm{RS}}\right]^{1/2},
    \label{eq:EinRing_eqGCJDF}
\end{equation}
where
\begin{subequations}
    \begin{align}
        & \mathfrak{A} = 12 M + 7 \alpha  \m^2 q^2 \rho_0 \mathcal{B}+6 \alpha  (1-4 k) q^2 \rho_0 \mathcal{B} \ln \xi -10 \alpha  k q^2 \rho_0 \mathcal{B}-20  \pi  \alpha  k q^2 \rho_0 \mathcal{B}+2 \alpha  (1-4 k) q^2 \rho_0 \mathcal{B} \ln \mathcal{B}\nonumber\\
        &\qquad+8 \m q^2 \rho_0 \mathcal{B}+5  \pi  \alpha  q^2 \rho_0 \mathcal{B}+2 \alpha  q^2 \rho_0 \mathcal{B}-2 q^2 \rho_0 \mathcal{B},\\
        & \mathfrak{B} = 24 M + 23 \alpha  \m^2 q^2 \rho_0 \mathcal{B}-6 \alpha  (8 \m+7) q^2 \rho_0 \mathcal{B} \ln \xi +70 \alpha  \m q^2 \rho_0 \mathcal{B}-40  \pi  \alpha  \m q^2 \rho_0 \mathcal{B}-2 \alpha  (8 \m+7) q^2 \rho_0 \mathcal{B} \ln \mathcal{B}\nonumber\\
        &\qquad+16 \m q^2 \rho_0 \mathcal{B}-35 \pi  \alpha  q^2 \rho_0 \mathcal{B}-14 \alpha  q^2 \rho_0 \mathcal{B}+14 q^2 \rho_0 \mathcal{B}.
    \end{align}
   \label{eq:mathfrakAB}
\end{subequations}
It is straightforward to check that in the absence of the dark fluid, this radius is estimated as
\begin{equation}
\theta_{\mathrm{ring}}^{\mathrm{Sch}} = \Biggl[\frac{2M D_\mathrm{S}}{D_\mathrm{R} D_\mathrm{RS}}+\frac{2\sqrt{M D_\mathrm{R} D_\mathrm{S}  (D_\mathrm{R} D_\mathrm{S} M+2 D_\mathrm{RS})}}{D_\mathrm{R}^2 D_\mathrm{RS}}\Biggr]^{1/2},
    \label{eq:EinRing_Sch}
\end{equation}
and hence, it depends only on the the distance from the source to the lens, in the case that $D_\mathrm{R}$ is fixed. To provide astrophysical justifications, recall from Subsect. \ref{subsec:constraints} that the distance from M87* is $D_\mathrm{R} \approx 16.8\,\mathrm{Mpc}$, whereas from Sgr A* it is $D_\mathrm{R} \approx 8.127 \,\mathrm{kpc}$. In Fig. \ref{fig:EinRing}, we have plotted the profile of the angular radius of the Einstein ring for a fixed $\alpha$ and different values of the $\mathcal{B}$-parameter, respecting the allowed domain for these parameters, when the lenses are assumed to be M87* and Sgr A*. 
Comparing with the diagram in Fig. \ref{fig:deflection}(a), and based on the assumed distances for the source and receiver in Fig. \ref{fig:EinRing}, the angular radius of the Einstein ring is proportional to half of the deflection angle for each considered black hole. The profiles in the figure show that an increase in the $\mathcal{B}$-parameter leads to a larger angular radius of the ring. Despite the significant distance difference between M87* and Sgr A*, the variation in the ring's angular radius between these two black holes is minimal. However, for a given impact parameter, an increase in the $\mathcal{B}$-parameter results in a larger angular radius for M87* compared to Sgr A*, which is more evident for smaller impact parameters as highlighted in the magnified segment of the diagram. This effect is attributed to the cosmological term in the GCJDF black hole, which becomes more prominent at larger distances. According to the results in Fig. \ref{fig:EinRing}, for large impact parameters, we can approximate $\theta_{\mathrm{ring}}^{\mathrm{Sch}}\approx 36.3\,\mathrm{\mu as}$, while $\theta_{\mathrm{ring}}^{\mathrm{GCJDF}}\approx 37.6284\,\mathrm{\mu as}$ for M87* and $\theta_{\mathrm{ring}}^{\mathrm{GCJDF}}\approx 37.6261\,\mathrm{\mu as}$ for Sgr A* with $\mathcal{B}=0.03$. These values fall within the detection capabilities of modern astronomical detectors. For instance, the EHT has an angular resolution of approximately $20\,\mathrm{\mu as}$ at 345 GHz \cite{kim_event_2020}. Additionally, the Gaia space observatory can resolve around $7$-$20\,\mathrm{\mu as}$ \cite{liu_gravitational_2017}. It is noteworthy that the impact of the dark fluid becomes particularly evident for sources near the black hole, where the discrepancy between the ring's radius of the GCJDF black hole and the Schwarzschild black hole is more noticeable. Consequently, Einstein rings are more detectable for more distant black holes when the light sources are positioned closer to them.

\section{Summary and conclusions}\label{sec:conclusions}

In this paper, we investigated the optical properties of a static spherically symmetric black hole within the context of an anisotropic generalized Chaplygin-like gas. We began by reviewing the characteristics of the CG and its generalized counterpart (GCG), incorporating anisotropic properties through Jacobi elliptic functions, resulting in the GCJG model. We demonstrated that the EoS for the GCJG depends on the modulus of the elliptic function, the factor $\alpha$, and the parameter $\mB$, which measures the prominence of the gas. Assuming this gas as the medium surrounding a static spherically symmetric black hole, termed the GCJDF, we derived a specific density profile aligned with the anisotropic energy-momentum tensor. Upon solving the Einstein equations for this black hole spacetime, we identified a radial-dependent cosmological term within the metric function, reducing to the Schwarzschild metric when $\mB = 0$. The analysis of energy conditions revealed that while the DEC is violated outside the Schwarzschild radius, the SEC is satisfied, highlighting the inflationary nature of the fluid. We explored light propagation in this spacetime by employing the Lagrangian framework for null geodesics in static spherically symmetric spacetimes, leading to the effective potential and the shadow parametrization of the black hole. Due to the non-asymptotic flatness of the black hole, we used a finite distance method to calculate the celestial coordinates and the shadow radius. Additionally, the energy emission rate and the deflection angle of light rays were computed. Our findings indicate that {decreasing} anisotropy in the dark fluid results in larger shadow radii, with the Schwarzschild black hole providing the lower limit for shadow size and deflection angle, and the upper limit for evaporation rate. While the dark fluid enhances gravitational lensing, its anisotropy increases lensing efficacy as more light rays are deflected, with fewer being captured by the black hole. Comparing our results with the EHT observations of M87* and Sgr A*, we found that optimal parameter domains are $0<\alpha\lesssim 0.1$ and $0\leq\mathcal{B}\lesssim 0.03$. Finally, we analyzed the sizes of Einstein rings for both black holes, concluding that for M87*, the ring size becomes more prominent with increasing $\mB$ due to the dominance of the cosmological term at larger distances. Our investigation into the optical properties of a black hole associated with a dark field can provide insights into the interplay between dark fluid anisotropy and light deflection, while constraints on the spacetime parameters validated by comparisons with the EHT observations, can help us in further explorations. Future work should aim to refine these models and extend the analysis to rotating black holes and other complex systems. These findings not only enhance our theoretical understanding but also pave the way for more precise astrophysical observations in the near future.

\section*{Acknowledgements}
M.F. is supported by Universidad Central de Chile through the project No. PDUCEN20240008.  
J.R.V. is partially supported by Centro de F\'isica Teórica de Valparaíso (CeFiTeV). G.A.P. was supported by CONAHCyT through the program {\it Estancias Posdoctorales por México 2023(1)}. M.C. work was partially supported by S.N.I.I. (CONAHCyT-M\'exico). The authors express their special thanks to the referees whose insightful comments enhanced the presentation of the paper.

\appendix

\section{Derivation of the density profile in Eq. \eqref{densid}}\label{app:A}

The Einstein field equations \eqref{eqein1} and \eqref{eqein2} can be recast as
\begin{eqnarray}
    && \frac{f}{r^2}+\frac{f'}{r}-\frac{1}{r^2}=-\rho,\label{eq:A1}\\
    && \frac{f'}{r}+\frac{f''}{2}=p_t(\rho),\label{eq:A2}
\end{eqnarray}
where the expression for $p_t$ has been given in Eq. \eqref{prtang}. Differentiating Eq. \eqref{eq:A1} provides
\begin{equation}
-\frac{2f}{r^3}+\frac{f''}{r}+\frac{2}{r^3}=-\rho'.
    \label{eq:A3}
\end{equation}
On the other hand, from Eq. \eqref{eq:A2} we have
\begin{equation}
    \frac{f''}{r}=\frac{2}{r}\left[p_t(\rho)-\frac{f'}{r}\right].
    \label{eq:A4}
\end{equation}
Employing Eqs. \eqref{eq:A3} and \eqref{eq:A4}, we get
\begin{equation}
-\rho'=\frac{2}{r}\underbrace{\left[\frac{1}{r^2}-\frac{f}{r^2}-\frac{f'}{r}\right]}_{=\rho,\,\mathrm{from\, Eq.\,\eqref{eq:A1}}}+\frac{2p_t(\rho)}{r}=\frac{2}{r}\mathcal{F}(\rho),
    \label{eq:A5}
\end{equation}
in which $\mathcal{F}(\rho)=\rho+p_t(\rho)$, and by means of Eq. \eqref{prtang}, we have
\begin{equation}
\mathcal{F}(\rho) = \frac{3}{2\rho^\alpha}\left[
-B\m-(2\m'-1)\rho^{\alpha+1}+\frac{\m'}{B}\rho^{2(\alpha+1)}
\right].
    \label{eq:A6}
\end{equation}
This way, Eq. \eqref{eq:A5} leads to the following differential equation:
\begin{equation}
-2\frac{\ed r}{r}=\frac{\ed\rho}{\mathcal{F}(\rho)}.
    \label{eq:A7}
\end{equation}
To solve the above equation, let us propose the change of the variable $X \doteq \rho^{\alpha+1}$. Accordingly, the differential equation changes to
\begin{equation}
\frac{2}{3(\alpha+1)}\,\frac{\ed X}{-B\m-(2\m'-1)X+\frac{\m'}{B}X^2}=-2\frac{\ed r}{r}.
    \label{eq:A8}
\end{equation}
Assuming $B<0$, one needs to consider the absolute value of this parameter, and Eq. \eqref{eq:A8} reshapes to 
\begin{equation}
\frac{2}{3(\alpha+1)}\,\frac{\ed X}{|B|\m-(1-2\m)X-\frac{\m'}{|B|}X^2}=-2\frac{\ed r}{r}.
    \label{eq:A9}
\end{equation}
When $0<k<1$, one can integrate both sides of Eq. \eqref{eq:A9}, which yields
\begin{eqnarray}
    \frac{2}{3(\alpha+1)}\ln\left(\frac{\m'[X+|B|\,]}{|B|\m-\m'X}\right) &=& -2\ln{r}+\frac{2}{3(\alpha+1)}\ln{q^2}\nonumber\\
    \Longrightarrow\ln\left(\frac{\m'[X+|B|\,]}{|B|\m-\m'X}\right) &=& \ln{\left(\frac{q^2}{r^{3(\alpha+1)}}\right)}\nonumber\\
    \Longrightarrow \frac{\m'\left(X+|B|\right)}{|B|\m-\m'X} &=& \frac{q^2}{r^{3(\alpha+1)}}\nonumber\\
    \stackrel{\mathrm{change\,of\,variable}}{\xRightarrow{\hspace*{2cm}}}
    \frac{\m'\left(\rho^{\alpha+1}+|B|\right)}{|B|\m-\m'\rho^{\alpha+1}} &=& \frac{q^2}{r^{3(\alpha+1)}}\doteq y(r).
    \label{eq:A10}
\end{eqnarray}
By defining $B=-\rho_0^{\alpha+1}\mathcal{B}$ with $(\rho_0, \mathcal{B})>0$ to ensure the condition $B<0$, Eq. \eqref{eq:A10} yields
\begin{equation}
\frac{\m'\left({\rho}/{\rho_0}\right)^{\alpha+1}\left(1+\mathcal{B}\right)}{\m'\left({\rho}/{\rho_0}\right)^{\alpha+1}\left(1-\m \mathcal{B}\right)}=y(r).
    \label{eq;A11}
\end{equation}
This will result in the solution \eqref{densid}.

\bibliographystyle{ieeetr}
\bibliography{biblio_v1.bib}

\end{document}